\documentclass[aps,preprintnumbers,showpacs,showkeys,nofootinbib]{revtex4}
\usepackage{amsmath,amssymb,bm,graphicx}

\renewcommand\d{\partial}
\newcommand\grad{\bm{\nabla}}
\newcommand\+{\dagger}
\newcommand\<{\langle}
\renewcommand\>{\rangle}
\newcommand\x{{\bm{x}}}
\newcommand\y{{\bm{y}}}
\newcommand\p{{\bm{p}}}
\newcommand\q{{\bm{q}}}
\renewcommand\k{{\bm{k}}}
\newcommand\eb{\varepsilon_\mathrm{b}}
\newcommand\eFA{\varepsilon_{\mathrm{F}\!A}}
\newcommand\kFA{k_{\mathrm{F}\!A}}
\newcommand\kFB{k_{\mathrm{F}\!B}}
\newcommand\nF{n_{\mathrm{F}}}
\newcommand\Tc{T_{\mathrm{c}}}
\newcommand\0{{\bm{0}}}
\newcommand\eff{{\mathrm{eff}}}
\newcommand\ind{{\mathrm{ind}}}
\renewcommand\O{\mathcal{O}}
\newcommand\m{\mathfrak{m}}
\renewcommand\j{\mathfrak{J}}

\begin{document}
\preprint{INT-PUB 08-46, MIT-CTP 3984}

\title{Induced $p$-wave superfluidity in two dimensions:\\
Brane world in cold atoms and nonrelativistic defect CFTs}
\author{Yusuke~Nishida}
\email{nishida@mit.edu}
\affiliation{Institute for Nuclear Theory, University of Washington,
             Seattle, Washington 98195-1550, USA}
\affiliation{Center for Theoretical Physics,
             Massachusetts Institute of Technology,
             Cambridge, Massachusetts 02139, USA}

\begin{abstract}
 We propose to use a two-species Fermi gas with the interspecies
 $s$-wave Feshbach resonance to realize $p$-wave superfluidity in two
 dimensions.  By confining one species of fermions in a two-dimensional
 plane immersed in the background three-dimensional Fermi sea of the
 other species, an attractive interaction is induced between
 two-dimensional fermions.  We compute the pairing gap in the
 weak-coupling regime and show that it has the symmetry of $p_x{+}ip_y$.
 Because the magnitude of the pairing gap increases toward the unitarity
 limit, it is possible that the critical temperature for the
 $p_x{+}ip_y$-wave superfluidity becomes within experimental reach.  The
 resulting system has a potential application to topological quantum
 computation using vortices with non-Abelian statistics.  We also
 discuss aspects of our system in the unitarity limit as a
 ``nonrelativistic defect conformal field theory (CFT)''.  The reduced
 Schr\"odinger algebra, operator-state correspondence, scaling
 dimensions of composite operators, and operator product expansions are
 investigated.
\end{abstract}

\date{October 2008}

\maketitle

\section{Introduction}
Experiments using ultracold atomic gases have achieved great success in
realizing a new type of fermionic superfluids.  By arbitrarily varying
the strength of interaction via the Feshbach resonance, the
weakly-interacting BCS superfluid, the strongly-interacting unitary
Fermi gas, and the Bose-Einstein condensate of tightly-bound molecules
have been observed and extensively
studied~\cite{Ketterle:2008,review_theory}.  So far, the fermionic
superfluids in atomic gases have been limited to $s$-wave pairings
between different spin states.  Therefore the realization of $p$-wave
superfluids in spin-polarized Fermi gases is a natural next goal in the
cold atom community.  In particular, a ``weakly-paired''
$p_x{+}ip_y$-wave superfluid in two dimensions is of special interest
because its vortices support zero-energy Majorana fermions and exhibit
non-Abelian statistics~\cite{Read:2000}.  As a practical application,
it has been proposed to use such a system as a platform for topological
quantum computation~\cite{Tewari:2007}.

Most theoretical studies regarding the $p$-wave superfluids in atomic
gases assume the availability of $p$-wave Feshbach
resonances~\cite{Tewari:2007,Bohn:2000,Ho:2005,Ohashi:2005,Gurarie:2005,Botelho:2005,Cheng:2005,Iskin:2006}
(for alternative mechanisms, see
Refs.~\cite{You:1999,Efremov:2000,Efremov:2002,Gaudio:2005,Bulgac:2006gh,Zhang:2008}).
However, experimental studies showed that the $p$-wave Feshbach
molecules are unstable due to atom-molecule and molecule-molecule
inelastic collisions with their lifetimes up to 20
ms~\cite{Regal:2003,Zhang:2004,Schunck:2005,Gunter:2005,Gaebler:2007,Fuchs:2008,Jin:2008,Inada:2008}.
This is in contrast to the long-lived $s$-wave Feshbach molecules where
the inelastic collisions are suppressed due to the Pauli exclusion
principle~\cite{Petrov:2004}.  Because the decay rate of the $p$-wave
Feshbach molecules is comparable to the interaction energy scale, the
$p$-wave superfluid without additional mechanism to suppress the
inelastic collisions will not reach its equilibrium before it
decays~\cite{Levinsen:2007,Jona-Lasinio:2008}.

In this paper, we propose a novel approach to realize the $p$-wave
superfluidity in two dimensions, without assuming the $p$-wave Feshbach
resonance.  The idea is to utilize a two-species Fermi gas (fermion
atomic species $A$ and $B$) with the interspecies $s$-wave Feshbach
resonance in 2D-3D mixed dimensions~\cite{Nishida:2008kr}.  Here $A$
atoms are confined in a two-dimensional plane (2D) by means of a strong
optical trap, while $B$ atoms are free from the confinement and hence in
the three-dimensional space (3D).  It has been shown that the
interspecies short-range interaction between $A$ and $B$ atoms is
characterized by a single parameter, the effective scattering length
$a_\eff$, whose value is arbitrarily tunable by the interspecies
$s$-wave Feshbach resonance~\cite{Nishida:2008kr}.  The system under
consideration can be set up in experiments with the use of the recently
observed quantum degenerate Fermi-Fermi mixture of ${}^6\mathrm{Li}$ and
${}^{40}\mathrm{K}$ atoms and their interspecies $s$-wave Feshbach
resonances~\cite{Taglieber:2008,Wille:2008}.

In such a system, we will show that the background 3D Fermi sea of $B$
atoms induces an attractive interaction between $A$ atoms in 2D.
Because $A$ atoms are identical fermions, the dominant pairing takes
place in the $p$-wave channel.   We will compute the pairing gap in the
controllable weak-coupling regime and show that it has the symmetry of
$p_x{+}ip_y$.  Because the magnitude of the pairing gap increases toward
the unitarity limit $|a_\eff|\to\infty$, the critical temperature for
the $p_x{+}ip_y$-wave superfluidity is expected to become within
experimental reach.  As it is mentioned above, the resulting system has
a potential application to topological quantum computation using
vortices with non-Abelian statistics~\cite{Read:2000,Tewari:2007}.

This paper is organized as follows.  In Sec.~\ref{sec:mixture}, we
describe the two-species Fermi gas in the 2D-3D mixed dimensions.  In
particular, we give its field-theoretical formulation in a detailed way
because such a system may not be familiar to the cold atom community.
Then in Sec.~\ref{sec:pairing}, we compute the induced interaction
between two-dimensional fermions, the pairing gap, and its symmetry in
the weak-coupling regime where we can perform the controlled
perturbative analysis.  Finally, summary and discussions are given in
Sec.~\ref{sec:summary} and here a very interesting analogy of the system
investigated in this paper with the brane-world model of the universe is
pointed out.  Two additional materials are presented in Appendices.  The
absence of the interspecies pairing at weak coupling is shown in the
Appendix~\ref{sec:absence}.  In the Appendix~\ref{sec:nrcft}, we discuss
aspects of our system in the unitarity limit as a {\em nonrelativistic
defect conformal field theory\/}.  We derive the reduced
Schr\"odinger algebra and the operator-state correspondence in general
nonrelativistic defect conformal field theories.  We also study scaling
dimensions of few-body composite operators and operator product
expansions in our 2D-3D mixed dimensions.  In particular, critical mass
ratios for Efimov bound states are obtained.

\section{Two-species Fermi gas in 2D-3D mixed dimensions
\label{sec:mixture}} 

\subsection{Field theoretical formulation}
The two-species Fermi gas in the 2D-3D mixed dimensions is described by
the following action (here and below $\hbar=1$ and $k_\mathrm{B}=1$):
\begin{equation}\label{eq:action}
 \begin{split}
  S &= \int\!dt\int\!d\x\,\psi_A^\+(t,\x)
  \left(i\d_t+\frac{\grad_{\!\x}^2}{2m_A}+\mu_A\right)\psi_A(t,\x) \\ 
  &\quad + \int\!dt\int\!d\x\!\int\!dz\,\psi_B^\+(t,\x,z)
  \left(i\d_t+\frac{\grad_{\!\x}^2+\nabla_{\!z}^2}{2m_B}
  +\mu_B\right)\psi_B(t,\x,z) \\
  &\quad + g_0\int\!dt\int\!d\x\,
  \psi_A^\+(t,\x)\psi_B^\+(t,\x,0)\psi_B(t,\x,0)\psi_A(t,\x).
 \end{split}
\end{equation}
Here $\x=(x,y)$ is a two-dimensional coordinate and $(\x,z)$ is a
three-dimensional coordinate.  $\psi_A(t,\x)$ is a fermionic field
describing $A$ atoms confined in a two-dimensional plane located at
$z=0$ and $\psi_B(t,\x,z)$ is another fermionic field describing $B$
atoms in the three-dimensional bulk space.  $m_{A(B)}$ is the atomic
mass of $A(B)$ atoms and the density of each species $n_{A(B)}$ is
controlled by the chemical potential $\mu_{A(B)}$.  The interspecies
interaction is short-ranged and thus occurs only on the plane at $z=0$,
while $B$ atoms can propagate into the $z$-direction (``extra
dimension'') [see also Fig.~\ref{fig:vacuum}].

$g_0$ is a cutoff dependent bare coupling.  Because dimensions of the
fields are $[\psi_A]=1$ and $[\psi_B]=\frac32$ in units of momentum, the
dimension of the coupling becomes $[g_0]=-1$.  This implies that the
theory has a linear divergence as it is well known in the usual 3D
case.  However, as we will see below, the linear divergence can be
renormalized into $g_0$ and all physical observables can be expressed in
terms of the physical parameter, the effective scattering length
$a_\eff$.  We note that interactions between the same species of
fermions (without the $p$-wave Feshbach resonance) are generally weak
and can be neglected at low energies.

The bare propagator of $\psi_A$ field is
$\<T\,\psi_A(t,\x)\psi_A^\+(t',\x')\>_0$ where the expectation value is
evaluated with the noninteracting action.  Because of the translational
symmetry in the plane, its Fourier transform is given by the usual form:
\begin{equation}\label{eq:G_A}
 iG_A(p_0,\p) = \frac{i}{p_0-\frac{\p^2}{2m_A}+\mu_A+i\delta},
\end{equation}
where $p_0$ is the frequency and $\p=(p_x,p_y)$ is the two-dimensional
momentum.  Similarly the bare propagator of $\psi_B$ field is given by
$\<T\,\psi_B(t,\x,z)\psi_B^\+(t',\x',z')\>_0$.  We shall not perform its
full Fourier transformation because once the interaction between
$\psi_A$ and $\psi_B$ fields is turned on, the translational symmetry
along the $z$-direction is lost.  Instead it is convenient to employ the
following mixed representation:
\begin{equation}\label{eq:G_B}
 iG_B(p_0,\p;z-z') = i\int\!\frac{dp_z}{2\pi}
  \frac{e^{ip_z(z-z')}}{p_0-\frac{\p^2+p_z^{\,2}}{2m_B}+\mu_B+i\delta},
\end{equation}
where $p_z$ is the momentum conjugate to $z-z'$.  We will often use the
propagator where $z$ and $z'$ are fixed on the plane; $z=z'=0$.  In such
a case, we suppress the last argument in $G_B(p_0,\p;z-z')$ and denote
it simply as $G_B(p_0,\p)\equiv G_B(p_0,\p;0)$.  Hereafter we shall use
a shorthand notation $p=(p_0,\p)$.

\subsection{Two-particle scattering in vacuum \label{sec:scattering}}

\begin{figure}[tp]
 \includegraphics[width=0.8\textwidth,clip]{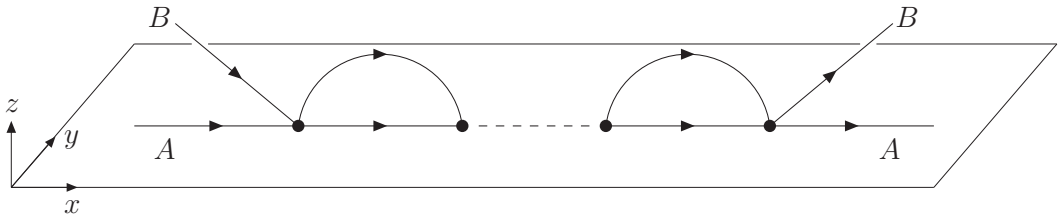}
 \caption{Two-particle scattering of $A$ and $B$ atoms.  The $A$ atom is
 confined in a plane located at $z=0$ while the $B$ atom can propagate
 into the $z$-direction (``extra dimension'').  The interspecies
 short-range interaction takes plane only on the plane.
 \label{fig:vacuum}}
\end{figure}

We first study the two-particle scattering in vacuum ($\mu_A=\mu_B=0$)
in order to relate the bare coupling $g_0$ with the effective scattering
length $a_\eff$.  The scattering process of $A$ and $B$ atoms is
schematically depicted in Fig.~\ref{fig:vacuum}.  By summing a geometric
series of Feynman diagrams, the scattering amplitude $\mathcal{A}(p)$ is
written as
\begin{equation}\label{eq:one-loop}
 \begin{split}
  [i\mathcal{A}(p)]^{-1} 
  &= \frac1{ig_0}-\int\!\frac{dk_0d\k}{(2\pi)^3}\,iG_A(p-k)\,iG_B(k) \\
  &= \frac1{ig_0} + i\int\!\frac{d\k}{(2\pi)^2}\frac{\sqrt{\frac{m_B}2}}
  {\sqrt{\frac{(\p-\k)^2}{2m_A}+\frac{\k^2}{2m_B}-p_0-i0^+}}. 
 \end{split}
\end{equation}
We can see that the $\k$ integration is ultraviolet divergent.  The
usual way to regulate the integral is to introduce a momentum cutoff
$|\k|<\Lambda_k$ and adjust the $\Lambda_k$-dependence of $g_0$ so that
the physics does not depend on $\Lambda_k$.  The integration over $\k$
leads to
\begin{equation}
 \mathcal{A}(p) = \frac1{\frac1{g_0} 
  - \frac{\sqrt{m_Bm_{AB}}}{2\pi}\left(\Lambda_k
  - \sqrt{\frac{m_{AB}}{M}\p^2-2m_{AB}p_0-i0^+}\right)},
\end{equation}
where $M=m_A+m_B$ is the total mass and $m_{AB}=\frac{m_Am_B}{m_A+m_B}$
is the reduced mass.  By introducing the effective scattering length
through
\begin{equation}\label{eq:a_eff}
 \frac1{g_0} - \frac{\sqrt{m_Bm_{AB}}}{2\pi}\Lambda_k
  = - \frac{\sqrt{m_Bm_{AB}}}{2\pi a_\eff},
\end{equation}
the scattering amplitude becomes cutoff-independent:
\begin{equation}\label{eq:T-matrix}
 \mathcal{A}(p) = \frac{2\pi}{\sqrt{m_Bm_{AB}}}
  \frac1{-\frac1{a_\eff}+\sqrt{\frac{m_{AB}}{M}\p^2-2m_{AB}p_0-i0^+}}.
\end{equation}

Now the interspecies interaction is solely characterized by the
effective scattering length $a_\eff$.  $a_\eff\to-0$ corresponds to the
weak attraction and $a_\eff\to+0$ corresponds to the strong attraction
just as in the usual 3D case.  $|a_\eff|\to\infty$ corresponds to the
unitarity limit where the scale-invariant interaction is achieved.  In
this limit, our theory (\ref{eq:action}) provides a novel type of
nonrelativistic conformal field theories.  Aspects of our system in the
unitarity limit as a nonrelativistic conformal field theory will be
elaborated in detail in the Appendix~\ref{sec:nrcft}.

When $a_\eff>0$, there exists a shallow two-body bound state composed of
$A$ and $B$ atoms.  Its binding energy $\eb$, defined to be positive, is
obtained as a pole of the scattering amplitude when the external
momentum $\p$ is zero:
\begin{equation}
 \mathcal{A}(-\eb,\0)^{-1}=0 \quad\Rightarrow\quad
 \eb = \frac1{2m_{AB}a_\eff^{\,2}}. 
\end{equation}
Thus our definition of $a_\eff$ in Eq.~(\ref{eq:T-matrix}) coincides
with that used in Ref.~\cite{Nishida:2008kr}.  The two-body resonance
$\eb\to0$ occurs at infinite effective scattering length
$a_\eff\to\infty$.

\subsection{Effective versus bare scattering lengths}

\begin{figure}[tp]\hfill
 \includegraphics[width=0.45\textwidth,clip]{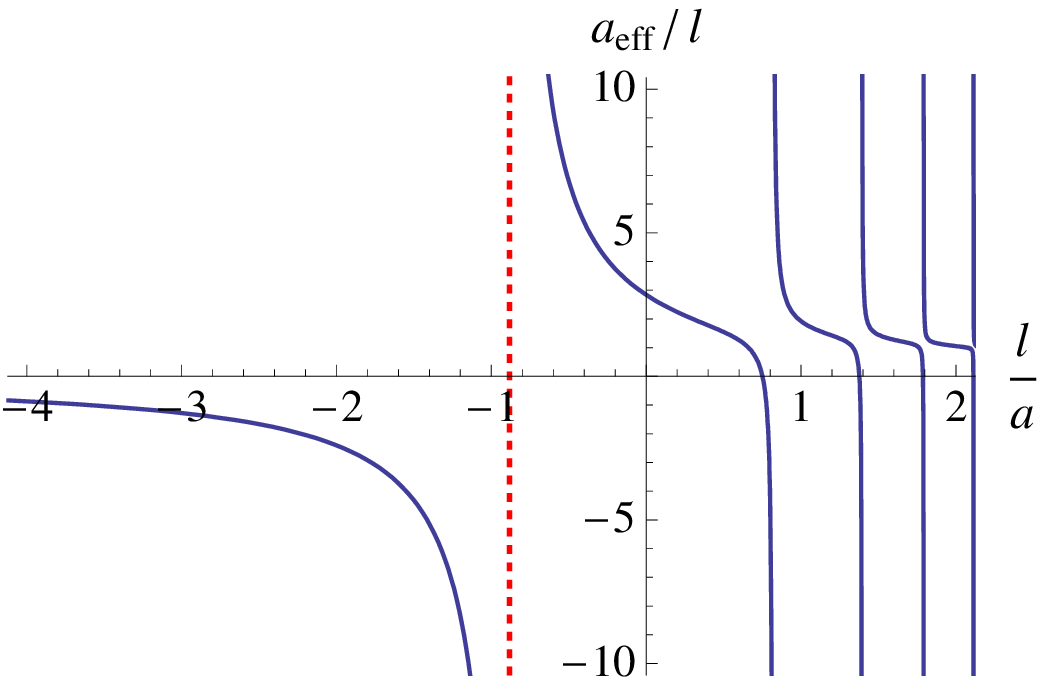}\hfill\hfill
 \includegraphics[width=0.45\textwidth,clip]{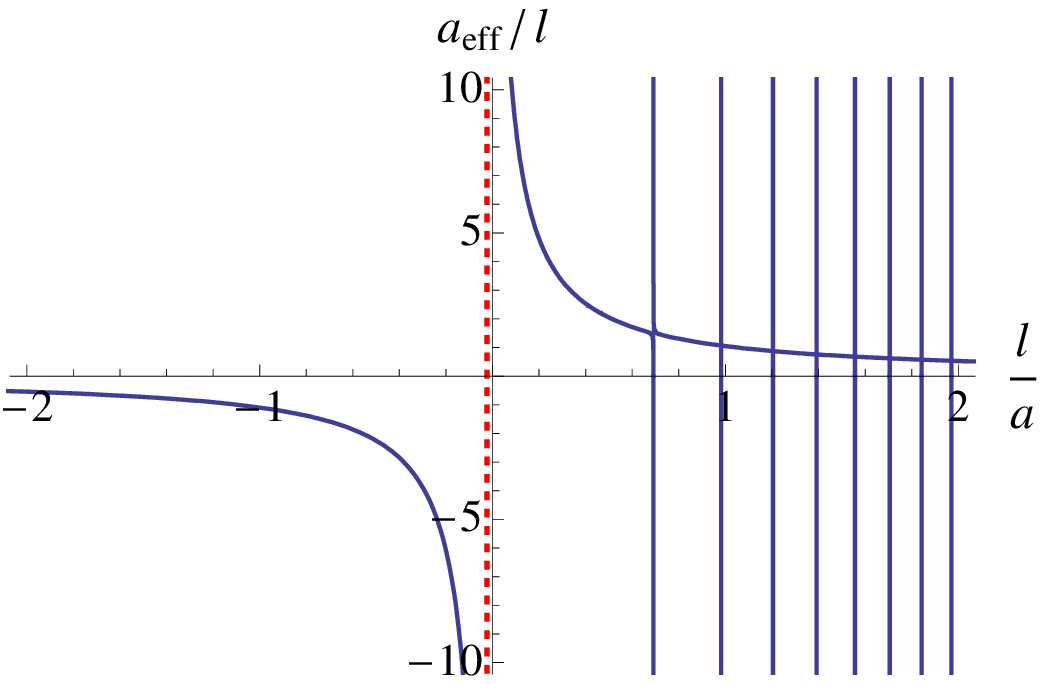}\hfill\hfill
 \caption{Effective scattering length $a_\eff/l$ as a function of the
 inverse bare scattering length $l/a$ for mass ratios $m_A/m_B=0.15$
 (left) and $m_A/m_B=6.67$ (right)~\cite{Nishida:2008kr}.  The vertical
 dotted line indicates the position of the broadest resonance.
 \label{fig:a_effective}}
\end{figure}

The effective scattering length $a_\eff$ in the 2D-3D mixed dimensions
depends on the bare scattering length $a$ in a free 3D space.  $a$ is
arbitrarily tunable by means of the interspecies $s$-wave Feshbach
resonance as a function of the magnetic field applied to the
system~\cite{Wille:2008}.  In Ref.~\cite{Nishida:2008kr}, the dependence
of $a_\eff$ on $a$ was determined when the $A$ atom is confined by a
one-dimensional harmonic potential with the oscillator frequency
$\omega_z$.  Fig.~\ref{fig:a_effective} shows $a_\eff/l$ plotted as a
function of $l/a$ with $l\equiv\sqrt{\frac1{m_A\omega_z}}$ being the
oscillator length~\cite{Nishida:2008kr}.  Here the mass ratios
$m_A/m_B=0.15$ and $6.67$ are chosen corresponding to the physical cases
of $A={}^6\mathrm{Li}$, $B={}^{40}\mathrm{K}$ and $A={}^{40}\mathrm{K}$,
$B={}^6\mathrm{Li}$, respectively.

We can see that the position of the resonance in the 2D-3D mixed
dimensions ($|a_\eff|=\infty$) is shifted from the free space resonance
($|a|=\infty$) to the negative bare scattering length.  It is
understandable that the $AB$ bound state can be formed with a weaker
attraction ($a<0$) because of the partial confinement of the $A$ atom.
The broadest resonance occurs at $l/a=-0.882$  
for $m_A/m_B=0.15$ and at $l/a=-0.0237$  
for $m_A/m_B=6.67$.  In addition to the broadest resonance, an infinite
number of confinement-induced resonances appears while they are
narrower~\cite{Peano:2005,Massignan:2006}.

Using one of these resonances, the effective scattering length can be
tuned to any desired value $-\infty<a_\eff^{-1}<\infty$ by simply
varying $a$ or $l$.  If the confinement length $l$ is much smaller than
any other length scales of the system such as $a_\eff$ and mean
interatomic distances at finite densities, we can neglect the motion of
$A$ atoms in the confinement $z$-direction.  Then the resulting system
becomes the two-species Fermi gas in the 2D-3D mixture universally
described by the action (\ref{eq:action}).

Ref.~\cite{Nishida:2008kr} also found that the many-body system near the
unitarity limit $|a_\eff|\to\infty$ is stable against the formation of
deep three-body bound states (Efimov effect) when the mass ratio is in
the range $0.0351<m_A/m_B<6.35$ (see also the
Appendix~\ref{sec:efimov}).  Therefore the combination of atomic
species, $A={}^6\mathrm{Li}$ and $B={}^{40}\mathrm{K}$ ($m_A/m_B=0.15$),
can be used to realize the stable 2D-3D mixed Fermi gas, while the
opposite combination, $A={}^{40}\mathrm{K}$ and $B={}^6\mathrm{Li}$
($m_A/m_B=6.67$), suffers the Efimov effect.  However, because the mass
ratio of the latter combination is just above the critical value, it may
be possible that such a system becomes metastable, for example, in an
optical lattice.  We also note that if either $A$ or $B$ atoms are
bosonic, the Efimov effect takes place for any mass
ratio~\cite{Nishida:2008kr}.  Thus for the stability of the many-body
system, fermion atomic species $A$ and $B$ are essential.

\subsection{Perturbation theory at finite density}
In the limit of weak attraction $a_\eff\to-0$, it is straightforward to
develop a perturbation theory at finite densities ($\mu_A,\,\mu_B>0$).
The propagator of $A$ atom is given by $iG_A(p)$ in Eq.~(\ref{eq:G_A})
and the propagator of $B$ atom is given by $iG_B(p;z)$ in
Eq.~(\ref{eq:G_B}).  From Eq.~(\ref{eq:T-matrix}), we find that each
interaction vertex carries a small coupling constant given by
$-\frac{2\pi ia_\eff}{\sqrt{m_Bm_{AB}}}$.

As one of applications of the perturbation theory, we compute the
density distribution of $B$ atoms in the weak-coupling limit
$a_\eff\to-0$.  Due to the lack of translational symmetry in the
$z$-direction, the density of $B$ atoms is no longer uniform.  The
density of $B$ atoms is given by
$\tilde n_B(|z|)=\<\psi_B^\+(t+0^+,\x,z)\psi_B(t,\x,z)\>$, which is a
function of $|z|$ because of the in-plane translational symmetry and the
symmetry under $z$-parity.  To the leading order in $a_\eff$, 
$\tilde n_B(|z|)$ is obtained as
\begin{equation}\label{eq:density}
 \begin{split}
  \tilde n_B(|z|) 
  &= -\int\!\frac{dp_0d\p}{(2\pi)^3}\,e^{ip_00^+}iG_B(p;0)
  - \int\!\frac{dp_0d\p}{(2\pi)^3}\,
  iG_B(p;z)\left(-i\Sigma_B\right)iG_B(p;-z) \\
  &= n_B -i \Sigma_B\int\!\frac{dp_0d\p dp_zdq_z}{(2\pi)^5}
  \frac{e^{ip_zz}}{p_0-\frac{\p^2+p_z^{\,2}}{2m_B}+\mu_B+i\delta}
  \frac{-e^{iq_zz}}{p_0-\frac{\p^2+q_z^{\,2}}{2m_B}+\mu_B+i\delta},
  \end{split}
\end{equation}
where $n_B=\frac{(2m_B\mu_B)^{3/2}}{6\pi^2}$ is the uniform density of
$B$ atoms in the noninteracting limit.  $\Sigma_B$ is a mean-field
self-energy proportional to the effective scattering length $a_\eff$ and
the density of $A$ atoms $n_A=\frac{(2m_A\mu_A)}{4\pi}$:
\begin{equation}
 \Sigma_B = -\frac{2\pi a_\eff}{\sqrt{m_Bm_{AB}}}
  \int\!\frac{dp_0d\p}{(2\pi)^3}\,e^{ip_00^+}iG_A(p)
  = \frac{2\pi a_\eff}{\sqrt{m_Bm_{AB}}}n_A <0.
\end{equation}
We note that the dimensions of $n_A$ and $n_B$ are different because
$n_A$ is the two-dimensional density while $n_B$ is the
three-dimensional density.  The Fermi momentum of each species is
defined through its density by $\kFA\equiv\left(4\pi n_A\right)^{1/2}$
and $\kFB\equiv\left(6\pi^2 n_B\right)^{1/3}$.

\begin{figure}[tp]
 \includegraphics[width=0.46\textwidth,clip]{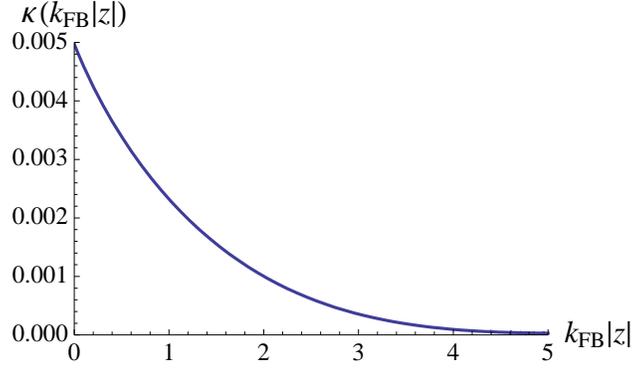}
 \caption{$\kappa(\kFB|z|)$ appearing in the density distribution of $B$
 atoms in Eq.~(\ref{eq:distribution}).  At $z=0$, we have
 $\kappa(0)=1/(64\pi)$.  \label{fig:density}}
\end{figure}

The integration over $p_0$ in Eq.~(\ref{eq:density}) results in the
following expression for the density distribution:
\begin{equation}\label{eq:distribution}
 \tilde n_B(|z|) = n_B + |a_\eff|\kFA^{\,2}\kFB^{\,2}
  \sqrt{\frac{m_B}{m_{AB}}}\,\kappa(\kFB|z|),
\end{equation}
where $\kappa(r)$ is a positive function given by
\begin{equation}
  \kappa(r) \equiv \int_0^1\!\frac{dp\,p}{2\pi}
  \int_{\sqrt{1-p^2}}^\infty\!\frac{dp_z}{2\pi}
  \int_0^{\sqrt{1-p^2}}\!\frac{dq_z}{2\pi}
  \frac{2\cos[(p_z-q_z)r]}{p_z^{\,2}-q_z^{\,2}}.
\end{equation}
$\kappa(\kFB|z|)$ is plotted in Fig.~\ref{fig:density} and monotonously
decreases as a function of $\kFB|z|$.  We can understand that $B$ atoms
are attracted to the 2D plane at $z=0$ because of their attractive
interaction with $A$ atoms confined in the plane.  The density of $B$
atoms away from the 2D plane approaches that in the noninteracting
limit; $\tilde n_B(|z|\to\infty)\to n_B$ because the interaction is suppressed
there.

\section{Induced interaction and $p$-wave pairing in two dimensions
\label{sec:pairing}}

\subsection{Induced interaction at weak coupling}
\begin{figure}[tp]
 \includegraphics[width=0.42\textwidth,clip]{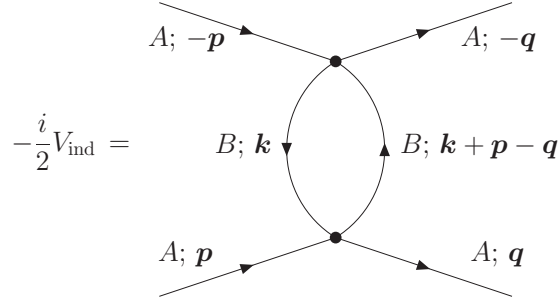}\hspace{1cm}
 \caption{Interaction between two $A$ atoms in 2D induced by the 3D
 Fermi sea of $B$ atoms.  \label{fig:medium}}
\end{figure}

Using the perturbation theory in the weak-coupling limit $a_\eff\to-0$,
we now determine the interaction between two $A$ atoms in 2D induced by
the existence of the 3D Fermi sea of $B$ atoms.  Because we are
interested in the intra-species pairing of $A$ atoms, we consider their
back-to-back scattering.  To the leading order in $a_\eff$, the induced
interaction between $A$ atoms $V_\ind$ is described by the Feynman
diagram depicted in Fig.~\ref{fig:medium}~\cite{Bulgac:2006gh}, which is
written as
\begin{equation}
 -\frac{i}2 V_\ind(p,q)
 = -\frac12\left(\frac{-2\pi ia_\eff}{\sqrt{m_Bm_{AB}}}\right)^2 
 \int\!\frac{dk_0d\k}{(2\pi)^3}\,iG_B(k+p-q)\,iG_B(k).
\end{equation}
The integration over $k_0$ leads to
\begin{equation}
 \begin{split}
  V_\ind(p,q) &= -\frac{\left(2\pi a_\eff\right)^2}{m_Bm_{AB}}
  \int\!\frac{d\k dk_zdk'_z}{(2\pi)^4} \\
  &\quad \times
  \left[\frac{\theta\!\left(\frac{(\k+\p-\q)^2+k_z^{\,2}}{2m_B}-\mu_B\right)
  \theta\!\left(\mu_B-\frac{\k^2+k_z^{\prime2}}{2m_B}\right)}
  {\frac{(\k+\p-\q)^2+k_z^{\,2}}{2m_B}
  -\frac{\k^2+k_z^{\prime2}}{2m_B}-p_0+q_0-i0^+}
  +\frac{\theta\!\left(\mu_B-\frac{(\k+\p-\q)^2+k_z^{\,2}}{2m_B}\right)
  \theta\!\left(\frac{\k^2+k_z^{\prime2}}{2m_B}-\mu_B\right)}
  {\frac{\k^2+k_z^{\prime2}}{2m_B}
  -\frac{(\k+\p-\q)^2+k_z^{\,2}}{2m_B}+p_0-q_0-i0^+}\right].
 \end{split}
\end{equation}

\begin{figure}[tp]
 \includegraphics[width=0.47\textwidth,clip]{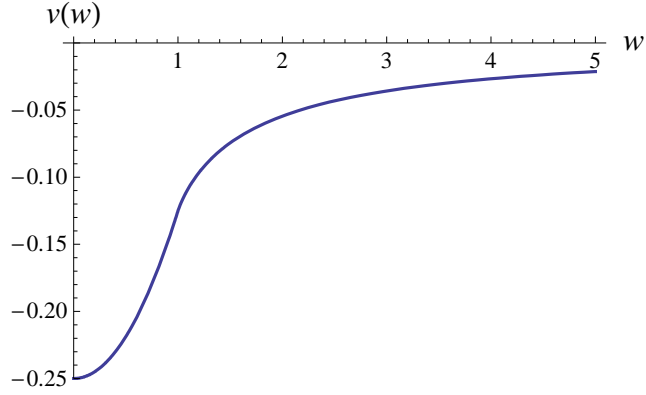}
 \caption{$v(w)$ appearing in the induced interaction in
 Eq.~(\ref{eq:induced}).  \label{fig:induced}}
\end{figure}

For the gap equation at weak coupling, we will need the induced
interaction in which both incoming and outgoing momenta are on the 2D
Fermi surface $|\p|=|\q|=\kFA$, and hence, $p_0=q_0=0$.  In such a
static limit, we can perform the remaining integrations analytically and
obtain
\begin{equation}\label{eq:induced}
 V_\ind(\p,\q) = \frac{2\pi a_\eff^{\,2}\kFB^{\,2}}{m_{AB}}
  \,v\!\left(\frac{|\p-\q|}{2\kFB}\right),
\end{equation}
where $v(w)$ is a continuous function given by
\begin{equation}
 v(w) \equiv
  \begin{cases}
   \displaystyle
   -\frac{2-w^2}8 & w<1 \medskip\\
   \displaystyle
   -\frac{\sqrt{w^2-1}+\left(2-w^2\right)\arcsin w^{-1}}{4\pi} & w>1.
  \end{cases}
\end{equation}
The nonanalyticity of $v(w)$ at $w=1$ is due to the sharp Fermi surface
of $B$ atoms.  The function $v(w)$ is plotted in Fig.~\ref{fig:induced}
and is negative everywhere indicating that the induced interaction
between $A$ atoms is attractive.  Thus an intra-species pairing in the
two-dimensional plane is expected to occur.  Because $A$ atoms are
identical fermions, the dominant pairing takes place in the $p$-wave
channel, as we will see below.

At this point, we should point out that the interspecies pairing between
$A$ and $B$ atoms is unlikely in our system (except deep in the BEC
regime $a_\eff\to+0$) because they live in different spatial dimensions.
$B$ atoms can always escape from the 2D plane in which $A$ atoms are
confined into the $z$-direction (``extra dimension'') and there the
interspecies interaction is turned off.  Actually, as we will show in
the Appendix~\ref{sec:absence}, the absence of the interspecies pairing
can be confirmed at weak coupling $a_\eff\to-0$.  Hereafter $B$ atoms
are treated as a background to induce the attraction between $A$ atoms
and we investigate the intra-species pairing of $A$ atoms in 2D.

\subsection{Gap equation}
Once the induced interaction between $A$ atoms
$V_\ind(\p,\q)$ is obtained, the pairing of $A$ atoms in 2D is described
by the BCS-type Hamiltonian:
\begin{equation}
 \begin{split}
  H_A &= \int\!\frac{d\p}{(2\pi)^2}
  \left(\frac{\p^2}{2m_A}-\mu_A\right)
  \tilde\psi_A^\+(\p)\tilde\psi_A(\p) \\
  &\quad + \frac12\int\frac{d\k d\p d\q}{(2\pi)^6}\,
  \tilde\psi_A^\+\!\left(\frac\k2+\q\right)
  \tilde\psi_A^\+\!\left(\frac\k2-\q\right)
  V_\ind(\p,\q)\,
  \tilde\psi_A\!\left(\frac\k2-\p\right)
  \tilde\psi_A\!\left(\frac\k2+\p\right),
 \end{split}
\end{equation}
where $\tilde\psi_A(\p)$ is the Fourier transform of $\psi_A(\x)$.  We
note the property $V_\ind(\p,\q)=V_\ind(\q,\p)$.  The pairing gap of $A$
atoms $\Delta_\p$ is defined to be
\begin{equation}
 (2\pi)^2\delta(\k)\Delta_\p
  = \int\!\frac{d\q}{(2\pi)^2}V_\ind(\p,\q)
 \left\<\tilde\psi_A\!\left(\frac{\k}2-\q\right)
 \tilde\psi_A\!\left(\frac{\k}2+\q\right)\right\>.
\end{equation}
Because of the Fermi statistics of $A$ atoms, the pairing gap has to
have an odd parity; $\Delta_{-\p}=-\Delta_{\p}$.  The standard
mean-field calculation leads to the following self-consistent gap
equation:
\begin{equation}\label{eq:gap_eq}
 \Delta_\p = -\int\!\frac{d\q}{(2\pi)^2}V_\ind(\p,\q)
  \frac{\Delta_\q}{2E_\q}\left[1-2\nF(E_\q)\right].
\end{equation}
Here $E_\p=\sqrt{\left(\frac{\p^2}{2m_A}-\mu_A\right)^2+|\Delta_\p|^2}$
is the quasiparticle energy and $\nF(E_\p)=1/\left(e^{E_\p/T}+1\right)$
is the Fermi-Dirac distribution function at temperature $T$.

The gap equation (\ref{eq:gap_eq}) is a nonlinear integral equation in
terms of the pairing gap $\Delta_\p$.  However, it becomes a linear
integral equation near the critical temperature $T\to\Tc$ because one
can set $\Delta_\p\to0$ in $E_\p$.  In such a case,
$\Delta_\p=e^{il\theta_{\hat\p}}\Delta^{(l)}$ with an odd integer $l$
being the orbital angular momentum solves the gap equation and the
critical temperature $\Tc$ is determined by the equation
\begin{equation}\label{eq:Tc_eq}
 1 = -N_AV_\ind^{(l)}
  \int_0^{\Lambda_\varepsilon}\!\frac{d\varepsilon}{\varepsilon}
  \tanh\!\left(\frac{\varepsilon}{2\Tc^{(l)}}\right).
\end{equation}
Here $\Lambda_\varepsilon$ is an energy cutoff and
$N_A\equiv\frac{m_A}{2\pi}$ is the density of states of $A$ atoms at the
Fermi surface.  $V_\ind^{(l)}=V_\ind^{(-l)}$ is the partial-wave
projection of the induced interaction given by
\begin{equation}\label{eq:projection}
 V_\ind^{(l)} = \frac{2\pi a_\eff^{\,2}\kFB^{\,2}}{m_{AB}}
  \int_0^\pi\!\frac{d\theta}{\pi}\cos(l\theta)\,
  v\!\left(\frac{\kFA}{\kFB}\sqrt{\frac{1-\cos\theta}2}\right)
  \equiv \frac{2\pi a_\eff^{\,2}\kFB^{\,2}}{m_{AB}}
  \,v^{(l)}\!\left(\frac{\kFA}{\kFB}\right).
\end{equation}
When the projected interaction is attractive $N_AV_\ind^{(l)}<0$, it is
easy to solve Eq.~(\ref{eq:Tc_eq}) and we find
\begin{equation}\label{eq:Tc}
 \frac{\Tc^{(l)}}{\eFA} \sim \exp\!\left(\frac1{N_AV_\ind^{(l)}}\right),
\end{equation}
where the energy cutoff is chosen to be the order of the Fermi energy of
$A$ atoms; $\Lambda_\varepsilon\sim\eFA=\frac{\kFA^{\,2}}{2m_A}$.
Because one can confirm that the induced attraction
(\ref{eq:projection}) is strongest in the $p$-wave channel, we have the
highest critical temperature for $|l|=1$;
$\Tc^{(1)}\gg\Tc^{(|l|\geq3)}$.  Therefore we can neglect the coupling
between different partial waves and concentrate on $p$-wave pairings.

\subsection{Pairing gap and its symmetry}
We now solve the gap equation for the $p$-wave pairing at zero
temperature $T=0$.  We parameterize the angle dependence of the pairing
gap as $\Delta_\p=f(\theta_{\hat\p})\Delta_f$, where
$\p=\kFA(\cos\theta_{\hat{\p}},\sin\theta_{\hat{\p}})$ and
$f(\theta_{\hat\p})=b_+e^{i\theta_{\hat\p}}+b_-e^{-i\theta_{\hat\p}}$
with $|b_+|^2+|b_-|^2=1$.  For example, $b_+=1$ and $b_-=0$ corresponds
to a $p_x{+}ip_y$-wave pairing and $b_+=b_-=1/\sqrt{2}$ corresponds to a
$p_x$-wave pairing.  Substituting $\Delta_\p=f(\theta_{\hat\p})\Delta_f$
into the gap equation (\ref{eq:gap_eq}) at $T=0$, we obtain
\begin{equation}
 \begin{split}
  1 &= -N_AV_\ind^{(1)}\int_0^{\Lambda_\varepsilon}\!d\varepsilon
  \int_0^{2\pi}\!\frac{d\theta_{\hat\q}}{2\pi}
  \frac{|f(\theta_{\hat\q})|^2}
  {\sqrt{\varepsilon^2+|f(\theta_{\hat\q})\Delta_f|^2}} \\
  &\simeq -N_AV_\ind^{(1)}\int_0^{2\pi}\!\frac{d\theta_{\hat\q}}{2\pi}
  |f(\theta_{\hat\q})|^2\ln\!\left(\frac{2\Lambda_\varepsilon}
  {|f(\theta_{\hat\q})\Delta_f|}\right).
 \end{split}
\end{equation}
Thus we find that the modulus of the pairing gap is given by
\begin{equation}\label{eq:modulus}
 \frac{|\Delta_f|}{\eFA} \sim \exp\!\left(\frac1{N_AV_\ind^{(1)}}
  -\int_0^{2\pi}\!\frac{d\theta}{2\pi}|f(\theta)|^2\ln|f(\theta)|\right).
\end{equation}

The angle dependence of the pairing gap $f(\theta)$ is determined so
that the ground state energy is minimized~\cite{Anderson:1961}.  Because
the gain of energy density due to the condensation is given by
\begin{equation}
 \<\mathcal{H}_A\> - \<\mathcal{H}_A\>\bigr|_{|\Delta_f|=0}
  = -\frac{N_A}{4}|\Delta_f|^2, 
\end{equation}
the ground state energy is minimized when $|\Delta_f|$ is maximized.
From Eq.~(\ref{eq:modulus}), we can show that the maximum $|\Delta_f|$
is achieved when $f(\theta)=e^{\pm i\theta}$ corresponding to the
$p_x{\pm}ip_y$-wave pairing.  Therefore the pairing gap realized in our
system becomes
\begin{equation}\label{eq:pairing}
 \frac{\Delta_\p}{\eFA} \sim e^{\pm i\theta_{\hat\p}}
  \exp\!\left(\frac1{N_AV_\ind^{(1)}}\right).
\end{equation}
We note that the pairing symmetry $p_x{\pm}ip_y$ is favored because of
the isotropy of the induced interaction; $V_\ind^{(1)}=V_\ind^{(-1)}$.
This is in contrast to the $p$-wave Feshbach resonance where the
interatomic interaction can be anisotropic due to the magnetic
dipole-dipole interaction~\cite{Ticknor:2004}.  In such a case, some
parameter-tunings are necessary to realize the $p_x{\pm}ip_y$-wave
pairing~\cite{Gurarie:2005,Cheng:2005}.

\subsection{Optimizing the pairing gap}
\begin{figure}[tp]
 \includegraphics[width=0.48\textwidth,clip]{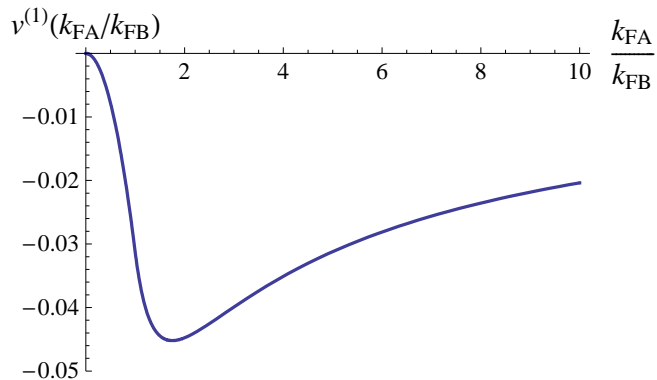}
 \caption{$v^{(1)}(\kFA/\kFB)$ appearing in the $p$-wave
 projection of the induced interaction in Eq.~(\ref{eq:projection}).
 \label{fig:fixed_akFB}}
\end{figure}

In order for the experimental realization of the proposed
$p_x{+}ip_y$-wave superfluidity, the critical temperature (\ref{eq:Tc})
and the magnitude of the pairing gap (\ref{eq:pairing}) have to be large
enough.  Thus we look for a condition in which
$\Tc^{(1)}\sim|\Delta_\p|\sim\eFA\exp\!\left[1/\left(N_AV_\ind^{(1)}\right)\right]$
is maximized.  From Eq.~(\ref{eq:projection}), $N_AV_\ind^{(1)}$ is
given by
\begin{equation}
 N_AV_\ind^{(1)} = \frac{m_A a_\eff^{\,2}\kFB^{\,2}}{m_{AB}}
  \,v^{(1)}\!\left(\frac{\kFA}{\kFB}\right)
\end{equation}
with the negative function $v^{(1)}(\kFA/\kFB)$ plotted in
Fig.~\ref{fig:fixed_akFB}.  Because the induced interaction in the
$p$-wave channel is attractive $N_AV_\ind^{(1)}<0$, one would like to
minimize $N_AV_\ind^{(1)}$.

One possible way to optimize the pairing gap is to control the densities
of $A$ and $B$ atoms~\cite{Bulgac:2006gh}.  Because our perturbative
calculation relies on the smallness of $|a_\eff\kFB|$, we fix
$a_\eff\kFB$ and vary the ratio in the two Fermi momenta $\kFA/\kFB$.
We find that the function $v^{(1)}(\kFA/\kFB)$ has an minimum
$v^{(1)}=-0.0452$  
at $\kFA/\kFB=1.75$ (see Fig.~\ref{fig:fixed_akFB}).  
Thus, within our perturbative calculation, the maximum pairing gap
becomes  
\begin{equation}
 \frac{\Delta_\p^{\mathrm{max}}}{\eFA} \sim e^{\pm i\theta_{\hat\p}}
  \exp\!\left(-\frac{22.1\,m_{AB}}{m_A a_\eff^{\,2}\kFB^{\,2}}\right).
\end{equation}

The pairing gap can be further enhanced by changing the mass ratio
$m_A/m_B$.  Because of the factor $m_{AB}/m_A$ in the exponent, the
larger mass of $A$ atoms in 2D increases the pairing gap.  For example,
the combination of atomic species $A={}^{40}\mathrm{K}$ and
$B={}^6\mathrm{Li}$ has $m_A/m_B=6.67$, and hence, the pairing gap
becomes  
\begin{equation}
 \frac{\Delta_\p^{\mathrm{Max}}}{\eFA} \sim e^{\pm i\theta_{\hat\p}}
  \exp\!\left(-\frac{3.29}{a_\eff^{\,2}\kFB^{\,2}}\right).
\end{equation}
Now the exponential factor is not hopelessly small.  If one could
extrapolate our perturbative result to $|a_\eff\kFB|\approx1$, we would
have
$\exp\!\left(-\frac{3.29}{a_\eff^{\,2}\kFB^{\,2}}\right)\approx0.04$.
Furthermore, in the unitarity limit $|a_\eff|\to\infty$, the pairing gap
is expected to be the same order as the Fermi energy;
$\Delta_\p\sim e^{\pm i\theta_{\hat\p}}\eFA$.  Therefore it is possible
that the critical temperature for the $p_x{+}ip_y$-wave superfluidity
becomes within experimental reach, in particular, near the unitarity
limit.

\subsection{Nonperturbative approaches near the unitarity limit
  \label{sec:epsilon}}
So far, we have performed the controlled perturbative analysis in the
weak-coupling regime $a_\eff\to-0$.  An important quantitative question
is how high the critical temperature $\Tc^{(1)}$ can be near the
unitarity limit $|a_\eff|\to\infty$.  Ideally one would like to answer
this question by employing quantum Monte Carlo simulations while they
will suffer fermion sign problems because of the intrinsic asymmetry
between $A$ and $B$ atoms in our 2D-3D mixture.  Instead it is possible
to estimate $\Tc^{(1)}$ by using nonperturbative analytical methods such
as the $\epsilon$
expansion~\cite{Nishida:2006br,Nishida:2006eu,Nishida:2006rp,Nishida:2006wk}
and the $1/N$ expansion~\cite{Nikolic:2007,Veillette:2007}.

The application of the $1/N$ expansion technique to our 2D-3D mixture is
straightforward.  We generalize the two-species Fermi gas in
Eq.~(\ref{eq:action}) to a $(2N)$-species Fermi gas in which $N$ species
live in 2D while the other $N$ species live in 3D.  The interaction
among them occurs on the 2D plane and is assumed to be the
$\mathrm{Sp}(2N)$-symmetric form~\cite{Nikolic:2007,Veillette:2007}.
Then we utilize the small parameter $1/N\ll1$ to perform systematic
expansions.

Here we comment on the application of the $\epsilon$ expansion technique
to mixed-dimensional systems.  Suppose $A$ and $B$ atoms live in $d_A$-
and $d_B$-dimensional spaces, respectively, where the former space is a
subset of the latter space with $d_A\leq d_B$.  Such a system is
described by the action analogous to Eq.~(\ref{eq:action}).  Now the
dimensions of the fields change to $[\psi_A]=d_A/2$ and $[\psi_B]=d_B/2$
in units of momentum, and thus, the dimension of the coupling becomes
$[g_0]=2-d_B$.  As far as $[g_0]>-2$ is satisfied, the theory is
renormalizable~\cite{Nishida:2006eu}.  We note that $[g_0]$ depends only
on the bulk spatial dimension $d_B$ indicating that $d_B$ plays a
central role in the $\epsilon$ expansion.  In the general combination of
the spatial dimensions, one can study the two-particle scattering in
vacuum as it was done in Sec.~\ref{sec:scattering}.  Using the
dimensional regularization, the scattering amplitude $\mathcal{A}(p)$ at
the scale-invariant unitarity point is found to be
\begin{equation}
 \mathcal{A}(p) = - \frac{\left(\frac{m_B}{m_{AB}}\right)^{d_A/2}
  \left(\frac{2\pi}{m_B}\right)^{d_B/2}}
  {\Gamma\!\left(1-\frac{d_B}2\right)
  \left(\frac{\p^2}{2M}-p_0-i0^+\right)^{d_B/2-1}},  
\end{equation}
where $\p$ is the $d_A$-dimensional momentum.  We can see that the
scattering amplitude vanishes in the limits of $d_B\to4$ and $d_B\to2$
indicating that these two spacial dimensions correspond to
noninteracting limits.  Accordingly we can develop systematic expansions
in terms of $\epsilon=4-d_B\ll1$ and $\bar\epsilon=d_B-2\ll1$ around
those special
dimensions~\cite{Nishida:2006br,Nishida:2006eu,Nishida:2006rp,Nishida:2006wk}.
We note that the other spatial dimension $d_A(\leq d_B)$ is arbitrary in
this approach.

The estimation of $\Tc^{(1)}$ near the unitarity limit using the above
nonperturbative approaches will be left for future works.

\section{Summary and discussions \label{sec:summary}}
In this paper, we presented theoretical prospects to realize the
$p$-wave superfluidity in two dimensions by using a two-species Fermi gas
(fermion atomic species $A$ and $B$) with the interspecies $s$-wave
Feshbach resonance.  By confining $A$ atoms in a 2D plane immersed in
the background 3D Fermi sea of $B$ atoms, an attractive interaction is
induced between $A$ atoms.  Because $A$ atoms are identical fermions,
the dominant pairing takes place in the $p$-wave channel.  In the
weak-coupling regime $a_\eff\to-0$ where the controlled perturbative
analysis is available in terms of the effective scattering length, we
computed the pairing gap and showed that it has the symmetry of
$p_x{+}ip_y$.  Because the magnitude of the pairing gap increases toward
the unitarity limit $|a_\eff|\to\infty$, the critical temperature for
the $p_x{+}ip_y$-wave superfluidity is expected to become within
experimental reach.  As it is mentioned in the Introduction, the
resulting system has a potential application to topological quantum
computation using vortices with non-Abelian
statistics~\cite{Read:2000,Tewari:2007}.

\begin{figure}[tp]
 \includegraphics[width=0.48\textwidth,clip]{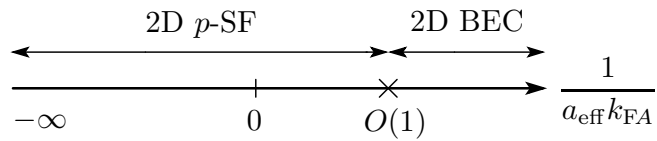}
 \caption{Proposed phase diagram of a two-species Fermi gas in the 2D-3D
 mixture as a function of the inverse effective scattering
 length $(a_\eff\kFA)^{-1}$ at zero temperature.  There is a quantum
 phase transition at $(a_\eff\kFA)^{-1}\simeq O(1)$ from the 2D-3D mixed
 Fermi gas with the 2D $p_x{+}ip_y$-wave superfluidity (2D $p$-SF) to
 the 2D Bose-Einstein condensation of localized $s$-wave molecules (2D
 BEC).  \label{fig:phase_diagram}}
\end{figure}

It is worthwhile to clarify what happens deep in the BEC regime
$a_\eff\to+0$ in our 2D-3D mixture.  In this limit, $A$ atoms in 2D
capture $B$ atoms to form tightly-bound molecules and the resulting
system consists of the molecules localized on the 2D plane plus excess
$A$ or $B$ atoms.  When the size of the molecules $\sim a_\eff$ becomes
smaller than the mean interatomic distance in 2D $\sim\kFA^{-1}$, the
molecules behave as two-dimensional bosons and therefore the ground
state will be a 2D Bose-Einstein condensate of the $s$-wave molecules.
Consequently, there has to be a quantum phase transition from the 2D-3D
mixed Fermi gas with the 2D $p_x{+}ip_y$-wave pairing
[$-\infty<(a_\eff\kFA)^{-1}\lesssim O(1)$] to the 2D Bose-Einstein
condensation of the $s$-wave molecules
[$O(1)\lesssim(a_\eff\kFA)^{-1}<+\infty$].  The proposed phase diagram
as a function of the inverse effective scattering length is shown in
Fig.~\ref{fig:phase_diagram}.  These two phases can be distinguished by
radio-frequency spectroscopy experiments.  In the 2D-3D mixed Fermi gas
with the 2D $p_x{+}ip_y$-wave pairing, $A$ atoms are fully gapped while
$B$ atoms remain gapless.  On the other hand, in the 2D Bose-Einstein
condensation of the $s$-wave molecules, both $A$ and $B$ atoms are fully
gapped.

Readers may wonder why we did not consider a system in which both $A$
and $B$ atoms are confined in a two-dimensional plane to realize the
induced $p$-wave superfluidity in 2D.  In this case, $A$ and $B$ atoms
always form bound molecules in 2D and thus the ground state of the
system tends to be an $s$-wave paired state.  In order to break the
interspecies $s$-wave pairing, one needs to weaken the interspecies
attraction with a large density imbalance introduced~\cite{He:2008}.
This would be a disadvantage in order to achieve a high critical
temperature for the $p$-wave superfluidity in 2D.

A remarkable aspect of our two-species fermions in the 2D-3D mixed
dimensions is that the system in the unitarity limit $|a_\eff|\to\infty$
is described by a nonrelativistic defect conformal field theory, which
is a novel class of quantum field theories that has not been paid
attention to so far.  We elaborated this aspect in detail in the
Appendix~\ref{sec:nrcft}.

Finally, it is very interesting to point out the analogy of the system
investigated in this paper with the brane-world model of the
universe.  In the brane-world scenario, the ordinary matter is
considered to be confined in a three-dimensional space (brane) embedded
in higher dimensions (bulk) where gravitons can
propagate~\cite{Maartens:2003tw}.  The gravitational force between
matters is induced by the exchange of the graviton.  Similarly, in our
system, the interaction between $A$ atoms confined in the 2D plane (``2D
brane'') is induced by the exchange of $B$ atoms in higher dimensions
(``3D bulk'').  Within this fascinating analogy, our two-species Fermi
gas in the 2D-3D mixed dimensions can be regarded as a brane world in
cold atoms!

\subsection*{Acknowledgments}
This work was motivated by the previous study~\cite{Nishida:2008kr}
performed in collaboration with S.~Tan to whom the author is grateful.
The author also thanks D.~T.~Son for introducing a notion of defect
conformal field theories to him.  This work was supported in part by
JSPS Postdoctoral Fellowship for Research Abroad and MIT Pappalardo
Fellowship in Physics.

\appendix

\section{Absence of interspecies pairing at weak coupling
\label{sec:absence}}
Here we show the absence of the interspecies pairing between $A$ and $B$
atoms in the 2D-3D mixed dimensions in the weak-coupling regime
$a_\eff\to-0$.  We consider the Nambu-Gor'kov-type propagator in the
$2\times2$ matrix form:
\begin{equation}
 i\mathcal G(t-t',\x-\x') =
  \begin{pmatrix}
   \<T\,\psi_A(t,\x)\psi_A^\+(t',\x')\> 
   & \<T\,\psi_A(t,\x)\psi_B(t',\x',0)\> \\
   \<T\,\psi_B^\+(t,\x,0)\psi_A^\+(t',\x')\>
   & \<T\,\psi_B^\+(t,\x,0)\psi_B(t',\x',0)\>
  \end{pmatrix}.
\end{equation}
In the mean-field approximation, the above propagator in the momentum
space becomes
\begin{equation}
 \tilde{\mathcal G}(p) =
  \begin{pmatrix}
   \frac{G_B^{-1}(-p)}{G_A^{-1}(p)G_B^{-1}(-p)+|\phi_0|^2}
   & \frac{\phi_0}{G_A^{-1}(p) G_B^{-1}(-p)+|\phi_0|^2} \\
   \frac{\phi_0^*}{G_A^{-1}(p)G_B^{-1}(-p)+|\phi_0|^2}
   & \frac{-G_A^{-1}(p)}{G_A^{-1}(p)G_B^{-1}(-p)+|\phi_0|^2}
  \end{pmatrix},
\end{equation}
where $\phi_0=\<g_0\,\psi_B(t,\x,0)\psi_A(t,\x)\>$ is a condensate
determined by the self-consistent gap equation:
\begin{equation}
 \begin{split}
  \frac{\phi_0}{g_0} &= -i\int\!\frac{dp_0d\p}{(2\pi)^3}
  \,\tilde{\mathcal G}_{12}(p) \\
  &= -\sqrt{\frac{m_B}2}\int\!\frac{dp_0 d\p}{(2\pi)^3}
  \frac{\phi_0}{\left(ip_0-\frac{\p^2}{2m_A}+\mu_A\right)
  \sqrt{ip_0+\frac{\p^2}{2m_B}-\mu_B}-\sqrt{\frac{m_B}2}|\phi_0|^2}.
 \end{split}
\end{equation}
In the last line, we analytically continued $p_0$ to the imaginary
frequency; $p_0\to ip_0$.  Introducing the effective scattering length
via Eq.~(\ref{eq:a_eff}), we obtain the following renormalized gap
equation:
\begin{equation}\label{eq:renormalized}
 -\frac{\sqrt{2m_{AB}}}{2\pi a_\eff}
  = \int\!\frac{dp_0 d\p}{(2\pi)^3}
  \left[\frac1{\left(ip_0-\frac{\p^2}{2m_A}\right)
   \sqrt{ip_0+\frac{\p^2}{2m_B}}}
   - \frac1{\left(ip_0-\frac{\p^2}{2m_A}+\mu_A\right)
   \sqrt{ip_0+\frac{\p^2}{2m_B}-\mu_B}
   -\sqrt{\frac{m_B}2}|\phi_0|^2}\right].
\end{equation}

For simplicity, we shall consider the equal masses $m_A=m_B=m$ and equal 
chemical potentials $\mu_A=\mu_B=\mu$ where the interspecies pairing is
guaranteed in the usual 3D case.  However, in the 2D-3D mixture, we can
see that the right-hand side of Eq.~(\ref{eq:renormalized}) does not
have any singularity around the Fermi surface, $p_0\sim0$ and
$\frac{\p^2}{2m}\sim\mu$, in the limit $|\phi_0|\to0$, and hence, the
integral is bounded from above.  This can be understood as an absence of
the Cooper instability because of the intrinsic ``mismatch'' between the
2D and 3D Fermi surfaces.  Therefore in the weak-coupling regime
$a_\eff\to-0$, the gap equation (\ref{eq:renormalized}) does not have a
nontrivial solution showing that there is no interspecies pairing
between $A$ and $B$ atoms.

\section{Aspects as a nonrelativistic defect conformal field theory
\label{sec:nrcft}}
As we mentioned in Sec.~\ref{sec:scattering}, two-species fermions in
the 2D-3D mixed dimensions in the unitarity limit $|a_\eff|\to\infty$
(at zero density and zero temperature) provide a novel type of
nonrelativistic conformal field theories (CFTs)~\footnote{As far as we
know, there are three basic ingredients to construct interacting
nonrelativistic CFTs; $1/R^2$-type interactions, zero-range interactions
at resonance~\cite{Mehen:1999nd}, and interactions due to fractional
statistics in two dimensions~\cite{Jackiw:1990mb}.  Combinations of
these interactions also work~\cite{Nishida:2007de,Nishida:2007mr}.  In
addition to those field-theoretical constructions, gravity dual
descriptions of different classes of nonrelativistic CFTs have been
recently proposed~\cite{Son:2008ye,Balasubramanian:2008dm}.  It would be
interesting to investigate gravity duals for nonrelativistic defect CFTs
imitating the situation studied in this paper.}.  In this system, the
three-dimensional translational, rotational, and Galilean symmetries in
the bulk space are broken to the two-dimensional symmetries while scale
and conformal invariance are preserved.  Regarding the two-dimensional
plane as a {\em defect\/} in the three-dimensional bulk space, our
system can be thought of a nonrelativistic counterpart of
defect/boundary CFTs~\cite{Cardy:1984bb,McAvity:1995zd}.  In
Ref.~\cite{Nishida:2008kr}, more classes of nonrelativistic defect CFTs
with zero-range and few-body resonant interactions have been proposed
and are summarized in Table~\ref{tab:mixtures}.  Here we discuss aspects
of our system as a nonrelativistic defect CFT (abbreviated as
\mbox{NRdCFT}).  First we derive the reduced Schr\"odinger algebra and
the operator-state correspondence in general nonrelativistic defect
CFTs.  Then we study scaling dimensions of few-body composite operators
and operator product expansions in our 2D-3D mixture.  In particular,
the critical mass ratios for the Efimov effect are obtained.

\begin{table}[tp]
 \caption{Summary of eight classes of nonrelativistic (defect) CFTs with
 zero-range and few-body resonant interactions proposed in
 Ref.~\cite{Nishida:2008kr}.  Two-species fermions in pure 3D are the
 well-known case of nonrelativistic
 CFT~\cite{Mehen:1999nd,Nishida:2007pj}.  In all cases below, the
 coupling of zero-range and few-body interaction term has the dimension
 $[g_0]=-1$ and is tuned to the resonance.  \label{tab:mixtures}}
 \begin{ruledtabular}
  \begin{tabular}{lll}
   \ Nonrelativistic (defect) CFT & Spatial configurations
   & Symmetries other than $H,\ D,\ C,\ \mathcal{M}$ \ \\\hline
   \ 2 species\ \ in\ \ pure 3D & $\x_A=\x_B=(x,y,x)$
   & $P_i,\ K_i,\ J_{ij}\quad\text{with}\quad i,j=x,y,z$ \\
   \ 2 species\ \ in\ \ 2D-3D mixture & $\x_A=(x,y)\quad\x_B=(x,y,z)$
   & $P_i,\ K_i,\ J_{ij}\quad\text{with}\quad i,j=x,y$ \\
   \ 2 species\ \ in\ \ 1D-3D mixture
   & $\x_A=(z)\quad\x_B=(x,y,z)$ & $P_z,\ K_z,\ J_{xy}$ \\
   \ 2 species\ \ in\ \ 2D-2D mixture
   & $\x_A=(x,z)\quad\x_B=(y,z)$ & $P_z,\ K_z$ \\
   \ 2 species\ \ in\ \ 1D-2D mixture
   & $\x_A=(z)\quad\x_B=(x,y)$ & $J_{xy}$ \\
   \ 3 species\ \ in\ \ 1D-1D-1D mixture
   & $\x_A=(x)\quad\x_B=(y)\quad\x_C=(z)$ & None \\
   \ 3 species\ \ in\ \ 1D$^2$-2D mixture
   & $\x_A=\x_B=(x)\quad\x_C=(x,y)$ & $P_x,\ K_x$ \\
   \ 4 species\ \ in\ \ pure 1D
   & $\x_A=\x_B=\x_C=\x_D=(x)$ & $P_x,\ K_x$
  \end{tabular}
 \end{ruledtabular} 
\end{table}

\subsection{Reduced Schr\"odinger algebra and operator-state correspondence}
Here we derive the reduced Schr\"odinger algebra and the operator-state
correspondence in general nonrelativistic defect CFTs. For definiteness,
we consider systems with two species of particles because the
generalization to more species is straightforward.  Define the mass
densities
\begin{equation}
 \begin{split}
  \m_A(\x_A) &= m_A\,\psi_A^\+(\x_A)\psi_A(\x_A) \\
  \m_B(\x_B) &= m_B\,\psi_B^\+(\x_B)\psi_B(\x_B)
 \end{split}
\end{equation}
and the momentum densities
\begin{equation}
 \begin{split}
  \bm{\j}_A(\x_A)
  &= -\frac{i}2\psi_A^\+(\x_A)\tensor{\grad}_{\!A}\psi_A(\x_A) \\
  \bm{\j}_B(\x_B) 
  &= -\frac{i}2\psi_B^\+(\x_B)\tensor{\grad}_{\!B}\psi_B(\x_B).
 \end{split}
\end{equation}
Here $\x_A$ ($\grad_{\!A}$) is a $d_A$-dimensional coordinate (derivative)
and $\x_B$ ($\grad_{\!B}$) is a $d_B$-dimensional coordinate
(derivative).  We assume that the intersection of the spaces in which
$A$ and $B$ particles live exists and includes the origin
$\x_A=\x_B=\0$.  For example, in our 2D-3D mixture, we have $\x_A=(x,y)$
and $\x_B=(x,y,z)$, while in general the $d_A$-dimensional space may not
be the subset of the $d_B$-dimensional space such as in the 2D-2D and
1D-2D mixtures in Table~\ref{tab:mixtures}.  We suppress the argument of
time when we denote the operators $\psi_A(t,\x)$ and $\psi_B(t,\x,z)$ at
$t=0$.

We consider commutation relations of the following set of operators in
general mixed dimensions: the Hamiltonian
\begin{equation}
 \begin{split}
  H &= \int\!d\x_A\frac{\grad_{\!A}\psi_A^\+(\x_A)
  \cdot\grad_{\!A}\psi_A(\x_A)}{2m_A}
  + \int\!d\x_B\frac{\grad_{\!B}\psi_B^\+(\x_B)
  \cdot\grad_{\!B}\psi_B(\x_B)}{2m_B} \\
  &\quad + \int\!d\x_A\!\int\!d\x_B\,\psi_A^\+(\x_A)\psi_B^\+(\x_B)
  V(\x_A,\x_B)\psi_B(\x_B)\psi_A(\x_A),
 \end{split}
\end{equation}
the dilatation operator
\begin{equation}\label{eq:dilatation}
 D = \int\!d\x_A\,\x_A\cdot\bm{\j}_A(\x_A)
  + \int\!d\x_B\,\x_B\cdot\bm{\j}_B(\x_B),
\end{equation}
and the special conformal operator
\begin{equation}\label{eq:conformal}
 C = \frac12\int\!d\x_A\,\x_A^{\,2}\,\m_A(\x_A)
  + \frac12\int\!d\x_B\,\x_B^{\,2}\,\m_B(\x_B).
\end{equation}
$D$ and $C$ are the generators of scale transformation
$\x_{A(B)}\to e^{\lambda}\x_{A(B)}$, $t\to e^{2\lambda}t$ and conformal
transformation $\x_{A(B)}\to\x_{A(B)}/(1+\lambda t)$, 
$t\to t/(1+\lambda t)$, respectively.  The commutation relation
\begin{equation}
 [D,\,C] = -2iC
\end{equation}
can be checked by a direct calculation.  By using the continuity
equation $[H,\,\m_{A}(x_{A})]=i\grad_{\!A}\cdot\bm{\j}_{A}(\x_{A})$
and the same with $A\to B$, we can show
\begin{equation}
 [H,\,C] = -iD.
\end{equation}
Finally, if the interparticle interaction $V(\x_A,\x_B)$ is scale
invariant (for example, $|\x_A-\x_B|^{-2}$-type interactions or
zero-range and infinite effective scattering length interactions
proposed in Ref.~\cite{Nishida:2008kr}), we obtain
\begin{equation}
 [D,\,H] = 2iH.
\end{equation}

If the system has unbroken translational, rotational, and Galilean
symmetries, the corresponding generators, namely, the momentum operators
\begin{equation}
 P_i = \int\!d\x_A\,\j_{Ai}(\x_A) + \int\!d\x_B\,\j_{Bi}(\x_B),
\end{equation}
the angular momentum operators
\begin{equation}
 J_{ij} = \int\!d\x_A\left[x_{Ai}\j_{Aj}(x_A)-x_{Aj}\j_{Ai}(x_A)\right]
  + \int\!d\x_B\left[x_{Bi}\j_{Bj}(x_B)-x_{Bj}\j_{Bi}(x_B)\right],
\end{equation}
and the Galilean boost operators
\begin{equation}
 K_i = \int\!d\x_A\,x_{Ai}\,\m_A(\x_A)
  + \int\!d\x_B\,x_{Bi}\,\m_B(\x_B),
\end{equation}
together with the above $H$, $D$, $C$, and the mass operator
\begin{equation}
 \mathcal{M} = \int\!d\x_A\m_A(\x_A) + \int\!d\x_B\m_B(\x_B)
\end{equation}
form the (reduced) Schr\"odinger algebra~\cite{Nishida:2007pj} (see
Table~\ref{tab:algebra}).
\begin{table}[tp]
 \caption{Full Schr\"odinger algebra in $d$ spatial dimensions
 $\mathrm{Sch}(d)$ taken from Ref.~\cite{Nishida:2007pj}.  The values of
 $[X,\,Y]$ are shown below.  The commutators of $\mathcal{M}$ and
 $J_{ij}$ with other operators are given by
 $[\mathcal{M},\,\text{any}]=[J_{ij},\,D]=[J_{ij},\,C]=[J_{ij},\,H]=0$,
 $[J_{ij},\,J_{kl}]=i(\delta_{ik}J_{jl}+\delta_{jl}J_{ik}-\delta_{il}J_{jk}-\delta_{jk}J_{il})$,
 $[J_{ij},\,P_k] =i(\delta_{ik}P_j-\delta_{jk}P_i)$, and
 $[J_{ij},\,K_k]=i(\delta_{ik}K_j-\delta_{jk}K_i)$ with
 $i,j=1,\ldots,d$.  \label{tab:algebra}}
 \begin{ruledtabular}
  \begin{tabular}{c|ccccc}
   \quad $X\,\backslash\,Y$ \quad\quad 
   & $P_j$ & $K_j$ & $D$ & $C$ & $H$ \quad\quad \\\hline  
   \quad $P_i$ \quad\quad 
   & $0$ & $-i\delta_{ij}\mathcal{M}$ & $-iP_i$ & $-iK_i$ & $0$ \quad\quad \\
   \quad $K_i$ \quad\quad 
   & $i\delta_{ij}\mathcal{M}$ & $0$ & $iK_i$ & $0$ & $iP_i$ \quad\quad \\
   \quad $D$ \quad\quad 
   & $iP_j$ & $-iK_j$ & $0$ & $-2iC$ & $2iH$ \quad\quad \\
   \quad $C$ \quad\quad 
   & $iK_j$ & $0$ & $2iC$ & $0$ & $iD$ \quad\quad \\
   \quad $H$ \quad\quad & $0$ 
       & $-iP_j$ & $-2iH$ & $-iD$ & $0$ \quad\quad
  \end{tabular}
 \end{ruledtabular}
\end{table}
Various classes of the reduced Schr\"odinger
algebra are possible depending on spatial configurations of defects as
shown in Table~\ref{tab:mixtures}.  For example, in our 2D-3D mixture,
there are planer translational, rotational, and Galilean symmetries
preserving the location of the 2D defect at $z=0$ and hence we can take
$P_i$, $K_i$, and $J_{ij}$ with $i,j=x,y$.  In some cases such as the
1D-1D-1D mixture with three species of particles, all translational,
rotational, and Galilean symmetries are broken by defects and thus only
$H$, $D$, $C$, and $\mathcal{M}$ form the reduced Schr\"odinger algebra.
We note that the symmetry transformations in the 2D-3D mixed dimensions
are not equivalent to those in the usual two dimensions although they
have the same Schr\"odinger algebra $\mathrm{Sch}(d=2)$.  This is
because the scale and conformal transformations generated by $D$ and $C$
in Eqs.~(\ref{eq:dilatation}) and (\ref{eq:conformal}) involve the
$z$-direction perpendicular to the 2D defect.

It is useful to introduce a notion of primary operators.  Consider a
local operator $\O(\x_{AB})$ composed of $\psi_A(\x_A)$ and
$\psi_B(\x_B)$ operators where $\x_{AB}=\x_A=\x_B$ is a coordinate on
the intersection of the $d_A$- and $d_B$-dimensional spaces including
the origin.  $\O(\x_{AB})$ is also called a defect operator because it
lives on the defect.  The local operator $\O$ is said to have a scaling
dimension $\Delta_\O$ and a mass $M_\O$ if it satisfies
\begin{equation}
 [D,\,\O(\0)] = i\Delta_\O\O(\0)
  \qquad\quad\text{and}\qquad\quad
  [\mathcal{M},\,\O(\0)] = M_\O\O(\0).
\end{equation}
Furthermore when $\O$ at origin commutes with $C$ and $K_j$ (if $K_j$
exists),
\begin{equation}
 [C,\,\O(\0)] = [K_j,\,\O(\0)] = 0,
\end{equation}
such a operator is called a primary operator.  Starting with the primary
operator $\O(\x_{AB})$, one can build up a tower of local operators by
repeatedly taking its commutators with $H$ and $P_j$ (if $P_j$
exists)~\cite{Nishida:2007pj}.  For example,
$[H,\,\O(\x_{AB})]=-i\d_t\O(\x_{AB})$ is a local operator having the
scaling dimension $\Delta_\O+2$ and
$[P_j,\,\O(\x_{AB})]=i\d_j\O(\x_{AB})$ is a local operator having the
scaling dimension $\Delta_{\O}+1$.

We are now ready to show the operator-state correspondence in
nonrelativistic defect CFTs.  Consider the state
\begin{equation}
 |\Psi_\O\> = e^{-H/\omega}\O^\+(\0)|0\>,
\end{equation}
where $\O$ is a primary operator. Then it is easy to show that
$|\Psi_\O\>$ is an energy eigenstate of the oscillator Hamiltonian
$H_\mathrm{osc}=H+\omega^2C$ with an energy eigenvalue
$\Delta_\O\omega$:
\begin{equation}\label{eq:correspondence}
 H_\mathrm{osc}|\Psi_\O\> 
  = \left(H+\omega^2C\right)e^{-H/\omega}\O^\+|0\>
  = e^{-H/\omega}\left(\omega^2C-i\omega D\right)\O^\+|0\>
  = \omega\Delta_\O|\Psi_\O\>.
\end{equation}
We note that the external potential term $\omega^2C$ in $H_\mathrm{osc}$
represents a $d_{A(B)}$-dimensional harmonic potential for $A(B)$
particles with $d_A$ and $d_B$ being different spatial dimensions in
general.  By further acting a raising operator
\begin{equation}
 L^\+=\frac{H}\omega-\omega\,C+iD
\end{equation}
to the primary state $|\Psi_\O\>$, we can generate a semi-infinite
ladder of energy eigenstates $(L^\+)^n|\Psi_\O\>$ with
$n=0,1,2,\ldots$~\cite{Nishida:2007pj}.  Their energy eigenvalues are
given by $\left(\Delta_\O+2n\right)\omega$ and can be interpreted as
excitations in the breathing mode~\cite{Werner:2006}.  If $P_j$ and
$K_j$ exist, one can make another raising operator
\begin{equation}
 Q_j^\+=\frac{P_j}{\sqrt{2\omega}}+i\sqrt{\frac{\omega}2}K_j,
\end{equation}
which generates energy eigenstates $(Q_j^\+)^n|\Psi_\O\>$ with energy
eigenvalues given by $\left(\Delta_\O+n\right)\omega$.  They correspond
to excitations in the center-of-mass motion.  The lowering operators
$L=\frac{H}\omega-\omega\,C-iD$ and
$Q_j=\frac{P_j}{\sqrt{2\omega}}-i\sqrt{\frac{\omega}2}K_j$ annihilate
the primary state; $L|\Psi_\O\>=0$ and $Q_j|\Psi_\O\>=0$.

Generalizations of other properties discussed in
Ref.~\cite{Nishida:2007pj} also hold in our nonrelativistic defect CFTs.
In particular, the two-point correlation function of the primary
operator $\O$ is determined up to an overall constant in terms of its
scaling dimension $\Delta_\O$ and its mass
$M_\O$~\cite{Henkel:1993sg,Nishida:2007pj}:
\begin{equation}\label{eq:correlation}
 \<T\,\O(t,\x_{AB})\O^\+(0,\0)\> \propto
  t^{-\Delta_\O}\exp\!\left(-iM_\O\frac{|\x_{AB}|^2}{2t}\right).
\end{equation}

\subsection{Composite operators and anomalous dimensions
\label{sec:efimov}}
We now turn to our specific nonrelativistic defect CFT, namely,
two-species fermions in the 2D-3D mixed dimensions (\ref{eq:action}) in
the unitarity limit $|a_\eff|\to\infty$.  Here we study various primary
operators and determine their scaling dimensions.  The simplest primary
operators are one-body operators $\psi_A(\x)$ and $\psi_B(\x,0)$ whose
scaling dimensions are trivially $\Delta_{\psi_A}=1$ and
$\Delta_{\psi_B}=3/2$, respectively.

A nontrivial primary operator is the two-body composite operator
\begin{equation}
 \phi(\x)\equiv\lim_{\y\to\x}|\y-\x|\psi_B(\y,0)\psi_A(\x).
\end{equation}
The presence of the prefactor $|\y-\x|$ guarantees that matrix elements
of the operator $\phi(\x)$ between two states in the Hilbert space are
finite.  Thus its scaling dimension becomes
\begin{equation}
 \Delta_{\phi} = \Delta_{\psi_A}+\Delta_{\psi_B}-1 = \frac32.  
\end{equation}
This result can be conformed by computing the two-point correlation
function of $\phi$ and comparing it with Eq.~(\ref{eq:correlation}):
\begin{equation}
 \<T\,\phi(t,\x)\phi^\+(0,\0)\>
  = -i\int\!\frac{dp_0d\p}{(2\pi)^3}\,e^{i\p\cdot\x-ip_0t}\mathcal{A}(p)
  \propto t^{-3/2}\exp\!\left(iM\frac{|\x|^2}{2t}\right).
\end{equation}
Here $\mathcal{A}(p)$ is the two-particle scattering amplitude given in
Eq.~(\ref{eq:T-matrix}) with $|a_\eff|\to\infty$.  The $\phi$ field can
be also interpreted as an auxiliary field that appears when we decompose
the four-Fermi interaction term in the action (\ref{eq:action}) using
the Hubbard-Stratonovich transformation;
$\phi(\x)=g_0\phi_B(\x,0)\psi_A(\x)$.  For the later use, we denote the
Fourier transform of the above $\phi$ propagator as
$iD(p)\equiv-i\mathcal{A}(p)$.

\subsubsection{$AAB$ three-body operators}
\begin{figure}[tp]
 \includegraphics[width=0.8\textwidth,clip]{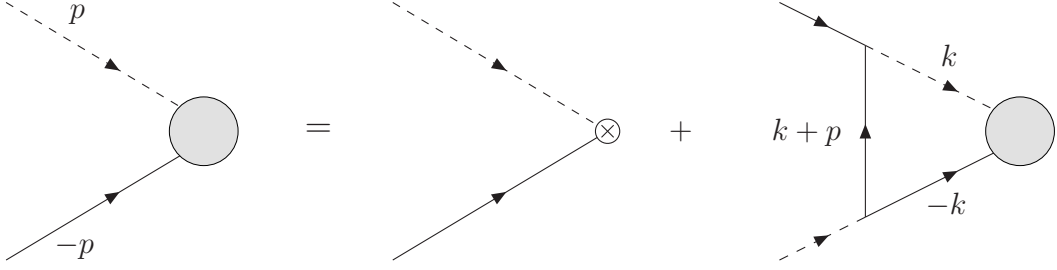}
 \caption{Feynman diagrams to renormalize the three-body composite
 operators $\phi\psi_{A(B)}$.  The solid lines are the propagators of
 $\psi_A$ and $\psi_B$ fields and the dotted lines are the propagators
 of $\phi$ field.  The shaded bulbs represent the vertex function
 $Z_{A(B)}(p)$.  \label{fig:3-body_vertex}}
\end{figure}

We then consider three-body composite operators.  A three-body operator
composed of two $A$ atoms and one $B$ atom with zero orbital angular
momentum $l=0$ is
\begin{equation}
 \O_{AAB}^{(l=0)}(\x) = Z_\Lambda^{-1}\phi(\x)\psi_A(\x),
\end{equation}
where $Z_\Lambda$ is a cutoff-dependent renormalization factor.  We
study the renormalization of the composite operator $\phi\psi_A$ by
evaluating its matrix element $\<0|\phi\psi_A(\x)|p,-p\>$.  Feynman
diagrams to renormalize $\phi\psi_A$ is depicted in
Fig.~\ref{fig:3-body_vertex}.  The vertex function $Z_A(p_0,\p)$ in
Fig.~\ref{fig:3-body_vertex} satisfies the following integral equation:
\begin{equation}
 \begin{split}
  Z_A(p_0,\p)
  &= 1 - i\int\!\frac{dk_0d\k}{(2\pi)^3}
  \,G_A(-k)G_B(k+p)D(k)Z_A(k_0,\k) \\
  &= 1 - \frac{\pi}{m_{AB}}\int\!\frac{d\k}{(2\pi)^2}
  \frac1{\sqrt{\frac{(\k+\p)^2}{2m_B}+\frac{\k^2}{2m_A}-p_0-i0^+}}
  \frac1{\sqrt{\frac{\k^2}{2M}+\frac{\k^2}{2m_A}-i0^+}}
  Z_A\!\left(-\frac{\k^2}{2m_A},\k\right),
 \end{split}
\end{equation}
where we used the analyticity of $Z_A(k_0,\k)$ on the upper half plane
of $k_0$.  The minus sign in front of the second term comes from the
Fermi statistics of $A$ atoms.  When we set $p_0=-\frac{\p^2}{2m_A}$,
$z_A(\p)\equiv Z_A\!\left(-\frac{\p^2}{2m_A},\p\right)$ satisfies
\begin{equation}\label{eq:integral_AAB}
 z_A(\p) = 1 - \frac{2\pi m_A}{m_{AB}}\int\!\frac{d\k}{(2\pi)^2}
 \frac1{\sqrt{\frac{m_A}{m_B}(\k+\p)^2+\k^2+\p^2}}
 \frac1{\sqrt{\frac{m_A}{M}\k^2+\k^2}}z_A(\k).
\end{equation}

Because of the scale invariance and in-plane rotational symmetry of the
system, we can assume the form of $z_A(\p)$ to be
\begin{equation}\label{eq:z_AAB}
 z_A(\p) = \frac1\chi\left(\frac{|\p|}\Lambda\right)^{\gamma+1},
\end{equation}
where $\Lambda$ is a momentum cutoff and $\chi$ is an unknown constant.
The renormalization factor becomes $Z_\Lambda\propto\Lambda^{-\gamma-1}$
and thus $\gamma+1$ is the anomalous dimension of the composite operator
$\phi\psi_A$.  (Here $\gamma$ and $\gamma_l$ appearing later are defined
so that they coincide with the definition of the scaling exponents used
in Ref.~\cite{Nishida:2008kr}.)  Once $\gamma$ is determined, the
scaling dimension of the renormalized composite operator
$\O_{AAB}^{(l=0)}$ is given by
\begin{equation}
 \Delta_{AAB}^{(l=0)} 
  = \Delta_{\phi}+\Delta_{\psi_A}+\gamma+1 = \frac72+\gamma.
\end{equation}
Substituting the form (\ref{eq:z_AAB}) into Eq.~(\ref{eq:integral_AAB}),
we obtain
\begin{equation}
 |\p|^{\gamma+1} = \chi\Lambda^{\gamma+1}
  - \frac{2\pi m_A}{m_{AB}\sqrt{\frac{m_A}{M}+1}}
  \int^\Lambda\!\!\frac{d\k}{(2\pi)^2}
  \frac{|\k|^\gamma}{\sqrt{\frac{m_A}{m_B}(\k+\p)^2+\k^2+\p^2}}.
\end{equation}
In order for the integral to be infrared finite,
$\mathrm{Re}\,(\gamma)>-2$ is necessary.  Also, in order to be able to
take the limit $\Lambda\to\infty$, $\mathrm{Re}\,(\gamma)<1$ is
required.  In the infinite cutoff limit $\Lambda\to\infty$, $\gamma$
satisfies the following equation:
\begin{equation}
 1 = -\frac{m_A}{m_{AB}\sqrt{\frac{m_A}{M}+1}}
  \int_0^\infty\!dk\int_0^\pi\!\frac{d\theta}\pi
  \frac{k^{\gamma+1}}{\sqrt{\left(\frac{m_A}{m_B}+1\right)k^2
  +\frac{2m_A}{m_B}k\cos\theta+\left(\frac{m_A}{m_B}+1\right)}},
\end{equation}
where $k=|\k|/|\p|$ and $\cos\theta=\hat{\k}\cdot\hat{\p}$.  Here the
integral is understood to be evaluated where it is convergent
$-2<\mathrm{Re}\,(\gamma)<-1$ and then analytically continued to an
arbitrary value of $\gamma$.

Similarly, for general orbital angular momentum $l$, we consider the
following three-body composite operator:
\begin{equation}
 \O_{AAB}^{(l)}(\x) = Z_\Lambda^{-1}\sum_{j=0}^lc_j
  \bigl(\d_x+i\d_y\bigr)^j\phi(\x)
  \bigl(\d_x+i\d_y\bigr)^{l-j}\psi_A(\x).
\end{equation}
In order for $\O_{AAB}^{(l)}$ to be a primary operator
($[K_i,\,\O_{AAB}^{(l)}]=0$), the coefficients $c_j$ have to be chosen
so that
\begin{equation}
 \sum_{j=0}^lc_i\left(p+\frac{M}{M+m_A}q\right)^j
  \left(-p+\frac{m_A}{M+m_A}q\right)^{l-j} \propto p^l
\end{equation}
being independent of the momentum $q$ conjugate to the center-of-mass
motion.  In the important case of $l=1$, we easily find
$c_1=-\frac{m_A}{M}c_0$.  If we denote the anomalous dimension of such a
composite operator as $\gamma_l+1-l$, it is straightforward to show that
$\gamma_l$ satisfies
\begin{equation}\label{eq:gamma_AAB}
 1 = -\frac{m_A}{m_{AB}\sqrt{\frac{m_A}{M}+1}}
  \int_0^\infty\!dk\int_0^\pi\!\frac{d\theta}{\pi}
  \frac{\cos(l\theta)\,k^{\gamma_l+1}}
  {\sqrt{\left(\frac{m_A}{m_B}+1\right)k^2
  +\frac{2m_A}{m_B}k\cos\theta+\left(\frac{m_A}{m_B}+1\right)}}.
\end{equation}
The integration over $k$ leads to the result shown in
Ref.~\cite{Nishida:2008kr}.  The scaling dimension of the renormalized
composite operator $\O_{AAB}^{(l)}$ is given by
\begin{equation}
 \Delta_{AAB}^{(l)} 
  = \Delta_{\phi}+\Delta_{\psi_A}+l+\left(\gamma_l+1-l\right)
  = \frac72+\gamma_l.
\end{equation}

\begin{figure}[tp]\hfill
 \includegraphics[width=0.46\textwidth,clip]{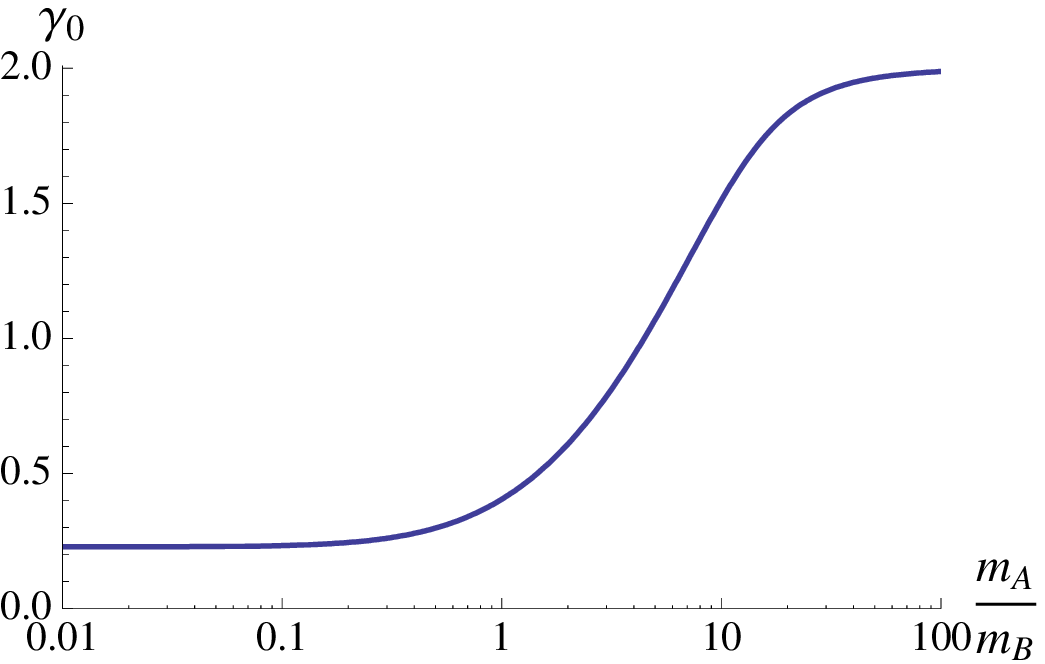}\hfill\hfill
 \includegraphics[width=0.46\textwidth,clip]{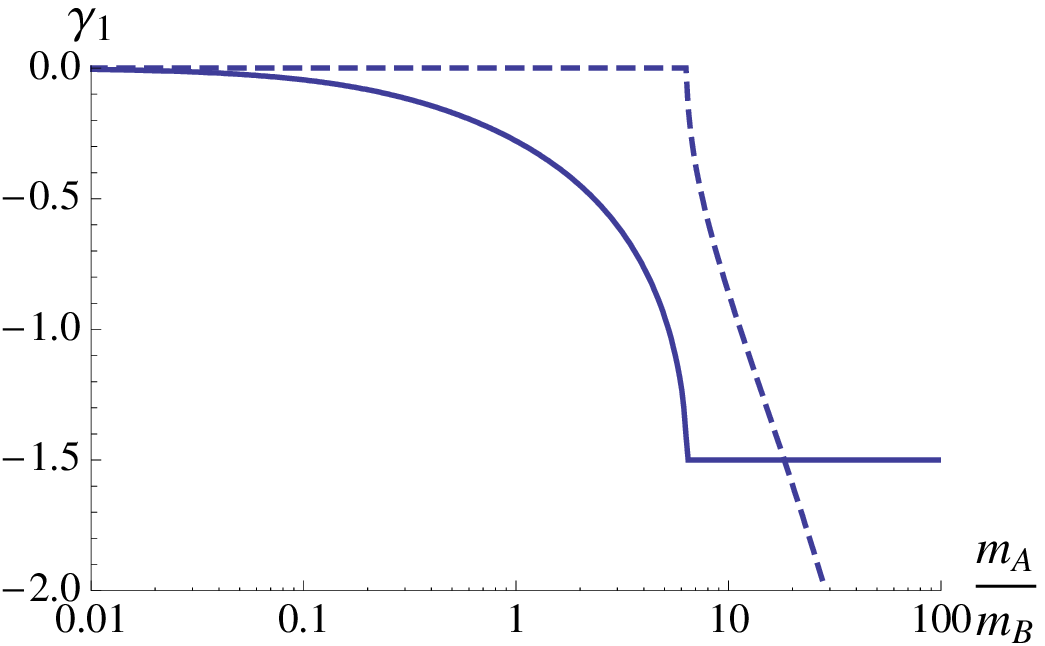}\hfill\hfill
 \caption{Anomalous dimensions $\gamma_l$ for $\O_{AAB}^{(l)}$ in the
 $s$-wave channel $l=0$ (left) and the $p$-wave channel $l=1$ (right) as
 functions of the mass ratio $m_A/m_B$~\cite{Nishida:2008kr}.  In the
 right panel, the real part of $\gamma_1$ (solid curve) and its
 imaginary part with a negative sign (dashed curve) are plotted.
 \label{fig:2d-fermion}}
\end{figure}

The anomalous dimensions $\gamma_l$ obtained by solving
Eq.~(\ref{eq:gamma_AAB}) in the $s$-wave channel $l=0$ and the $p$-wave
channel $l=1$ are plotted in Fig.~\ref{fig:2d-fermion} as functions of
the mass ratio $m_A/m_B$.  For $l=0$, $\gamma_0$ increases as $m_A/m_B$
is increased indicating the stronger effective repulsion in the
$s$-wave channel.  On the other hand, for $l=1$, $\gamma_1$ decreases
with increasing $m_A/m_B$ and eventually becomes complex as
$\gamma_1=-\frac32\pm i\,\mathrm{Im}\,(\gamma_1)$ when
$m_A/m_B>6.35111$~\cite{Nishida:2008kr}.  (For comparison, the
Born-Oppenheimer approximation predicts the critical mass ratio to be
$m_A/m_B\approx6.21791$.)  In this case, using the scaling dimension
$\Delta_{AAB}^{(l=1)}=2\pm i\,\mathrm{Im}\,(\gamma_1)$ and
Eq.~(\ref{eq:correlation}), the two-point correlation function of
$\O_{AAB}^{(l=1)}$ is found to behave as
\begin{equation}\label{eq:efimov}
 \begin{split}
  \int\!dtd\x\,e^{-i\p\cdot\x+ip_0t}\<T\,\O(t,\x)\O^\+(0,\0)\>
  &= \sum_{\pm} b_{\pm} \left[\frac{\p^2}{4m_A+2m_B}-p_0-i0^+\right]
  ^{\pm i\,\mathrm{Im}(\gamma_1)} \\
  &\propto \sin\!\left[\mathrm{Im}(\gamma_1)
  \ln\left(\frac{\p^2-(4m_A+2m_B)p_0-i0^+}{\Lambda^2}\right)+\varphi\right].
 \end{split}
\end{equation}
Now the full scale invariance is broken to a discrete scaling symmetry,
\begin{equation}\label{eq:discrete}
 \p\to e^{\pi/\mathrm{Im}(\gamma_1)}\,\p
  \qquad\quad\text{and}\qquad\quad
  p_0\to e^{2\pi/\mathrm{Im}(\gamma_1)}\,p_0, 
\end{equation}
which is a characteristic of a renormalization-group limit
cycle~\cite{Bedaque:1998kg}.  This implies the existence of an infinite
set of discrete bound states in the $p$-wave $AAB$ three-body system.
The energy eigenvalues form a geometric spectrum as
$E_{n+1}/E_n=e^{-2\pi/|\mathrm{Im}(\gamma_1)|}$ and they are known as
Efimov bound states in the usual 3D case~\cite{Efimov:1972}.  Because
the system develops deep three-body bound states, the corresponding
many-body system cannot be stable toward collapse.

We note that an interesting thing becomes possible in the range of mass
ratio $2.32780<m_A/m_B<6.35111$~\cite{Nishida:2008kr}.  Here the
anomalous dimension is $-\frac32<\gamma_1<-\frac12$ (see the right panel
in Fig.~\ref{fig:2d-fermion}) and thus the scaling dimension of the
three-body composite operator $\O_{AAB}^{(l=1)}$ becomes
$2<\Delta_{AAB}^{(l=1)}<3$.  Therefore a new three-body interaction term
\begin{equation}\label{eq:resonance}
 S_\text{3-body} = g_1\int\!dtd\x\,\O^\+(t,\x)\O(t,\x)
\end{equation}
becomes renormalizable because now the coupling has the dimension
$-2<[g_1]<0$~\cite{Nishida:2007mr}.  The action (\ref{eq:action}) with
$S_\text{3-body}$ added defines a new renormalizable theory.  In
particular, when $g_1$ is tuned to an $AAB$ three-body resonance, the
resulting system provides a novel nonrelativistic defect CFT describing
two-species fermions with both two-body ($AB$) and three-body ($AAB$)
resonances in the 2D-3D mixture~\cite{Nishida:2008kr}.

\subsubsection{$ABB$ three-body operators}
A three-body operator composed of one $A$ atom and two $B$ atoms with
zero orbital angular momentum $l=0$ is
\begin{equation}
 \O_{ABB}^{(l=0)}(\x) = Z_\Lambda^{-1}\phi(\x)\psi_B(\x,0), 
\end{equation}
where $Z_\Lambda$ is a cutoff-dependent renormalization factor.  We can
study the renormalization of the composite operator $\phi\psi_B$ by
evaluating its matrix element $\<0|\phi\psi_B(\x)|p,-p\>$.  Feynman
diagrams to renormalize $\phi\psi_B$ is depicted in
Fig.~\ref{fig:3-body_vertex}.  The vertex function $Z_B(p_0,\p)$ in
Fig.~\ref{fig:3-body_vertex} satisfies the following integral equation:
\begin{equation}
 \begin{split}
  Z_B(p_0,\p) &= 1 - i\int\!\frac{dk_0d\k}{(2\pi)^3}
  \,G_B(-k)G_A(k+p)D(k)Z_B(k_0,\k) \\
  &= 1 - \frac{\pi}{m_{AB}}\sqrt{\frac2{m_B}}
  \int\!\frac{d\k dk_z}{(2\pi)^3}
  \frac1{\frac{(\k+\p)^2}{2m_A}+\frac{\k^2+k_z^{\,2}}{2m_B}-p_0-i0^+}
  \frac1{\sqrt{\frac{\k^2}{2M}+\frac{\k^2+k_z^{\,2}}{2m_B}-i0^+}}
  Z_B\!\left(-\frac{\k^2+k_z^{\,2}}{2m_B},\k\right),
 \end{split}
\end{equation}
where we used the analyticity of $Z_B(k_0,\k)$ on the upper half plane
of $k_0$.  The minus sign in front of the second term comes from the
Fermi statistics of $B$ atoms.  When we set
$p_0=-\frac{\p^2+p_z^{\,2}}{2m_B}$,
$z_B(\p,p_z)\equiv Z_B\!\left(-\frac{\p^2+p_z^{\,2}}{2m_B},\p\right)$
satisfies 
\begin{equation}\label{eq:integral_ABB}
 \begin{split}
  z_B(\p,p_z) &= 1 - \frac{4\pi m_B}{m_{AB}}
  \int\!\frac{d\k dk_z}{(2\pi)^3}
  \frac1{\frac{m_B}{m_A}(\k+\p)^2+\k^2+k_z^{\,2}+\p^2+p_z^{\,2}}
  \frac1{\sqrt{\frac{m_B}{M}\k^2+\k^2+k_z^{\,2}}} z_B(\k,k_z).
 \end{split}
\end{equation}

Because of the scale invariance and in-plane rotational symmetry of the
system, we can assume the form of $z_B(\p,p_z)$ to be
\begin{equation}\label{eq:z_ABB}
 z_B(\p,p_z) = \left(\frac{|\p|}\Lambda\right)^{\gamma+1}
  S\!\left(\frac{|p_z|}{|\p|}\right),
\end{equation}
where $\Lambda$ is a momentum cutoff and $S(|p_z|/|\p|)$ is an unknown
function.  The renormalization factor becomes
$Z_\Lambda\propto\Lambda^{-\gamma-1}$ and thus $\gamma+1$ is the
anomalous dimension of the composite operator $\phi\psi_B$.  (Here
$\gamma$ and $\gamma_l$ appearing later are defined so that they
coincide with the definition of the scaling exponents used in
Ref.~\cite{Nishida:2008kr}.)  Once $\gamma$ is determined, the scaling
dimension of the renormalized composite operator $\O_{ABB}^{(l=0)}$ is
given by
\begin{equation}
 \Delta_{ABB}^{(l=0)} 
  = \Delta_{\phi}+\Delta_{\psi_B}+\gamma+1 = 4+\gamma.
\end{equation}
Substituting the form (\ref{eq:z_ABB}) into Eq.~(\ref{eq:integral_ABB}),
we obtain
\begin{equation}
 \begin{split}
  |\p|^{\gamma+1} S\!\left(\frac{|p_z|}{|\p|}\right)
  &= \Lambda^{\gamma+1} - \frac{4\pi m_B}{m_{AB}}
  \int^\Lambda\!\frac{d\k dk_z}{(2\pi)^3}
  \frac1{\frac{m_B}{m_A}(\k+\p)^2+\k^2+k_z^{\,2}+\p^2+p_z^{\,2}}
  \frac{|\k|^{\gamma+1}}{\sqrt{\frac{m_B}{M}\k^2+\k^2+k_z^{\,2}}}
  \,S\!\left(\frac{|k_z|}{|\k|}\right).
 \end{split}
\end{equation}
In order for the integral to be infrared finite,
$\mathrm{Re}\,(\gamma)>-3$ is necessary.  Also, in order to be able to
take the limit $\Lambda\to\infty$, $\mathrm{Re}\,(\gamma)<1$ is
required.  In the infinite cutoff limit $\Lambda\to\infty$, $\gamma$
satisfies the following integral equation:
\begin{equation}
 \begin{split}
  |\p|^{\gamma+1} S\!\left(\frac{|p_z|}{|\p|}\right)
  &= - \frac{4\pi m_B}{m_{AB}}
  \int_{-\infty}^\infty\!\frac{d\k dk_z}{(2\pi)^3}
  \frac1{\frac{m_B}{m_A}(\k+\p)^2+\k^2+k_z^{\,2}+\p^2+p_z^{\,2}}
  \frac{|\k|^{\gamma+1}}{\sqrt{\frac{m_B}{M}\k^2+\k^2+k_z^{\,2}}}
  \,S\!\left(\frac{|k_z|}{|\k|}\right).
 \end{split}
\end{equation}
Here the integral is understood to be evaluated where it is convergent
$-3<\mathrm{Re}\,(\gamma)<-1$ and then analytically continued to an
arbitrary value of $\gamma$.

Similarly, for general orbital angular momentum $l$, we consider the
following three-body composite operator:
\begin{equation}\label{eq:O_ABB}
 \O_{ABB}^{(l)}(\x) = Z_\Lambda^{-1}\sum_{j=0}^lc_j
  \bigl(\d_x+i\d_y\bigr)^j\phi(\x)
  \bigl(\d_x+i\d_y\bigr)^{l-j}\psi_B(\x).
\end{equation}
In order for $\O_{ABB}^{(l)}$ to be a primary operator
($[K_i,\,\O_{ABB}^{(l)}]=0$), the coefficients $c_j$ have to be chosen
so that
\begin{equation}
 \sum_{j=0}^lc_i\left(p+\frac{M}{M+m_B}q\right)^j
  \left(-p+\frac{m_B}{M+m_B}q\right)^{l-j} \propto p^l
\end{equation}
being independent of the momentum $q$ conjugate to the center-of-mass
motion.  In the important case of $l=1$, we easily find
$c_1=-\frac{m_B}{M}c_0$.  If we denote the anomalous dimension of such a
composite operator as $\gamma_l+1-l$, it is straightforward to show that
$\gamma_l$ satisfies
\begin{equation}\label{eq:gamma_ABB}
 \begin{split}
  |\p|^{\gamma_l+1} S_l\!\left(\frac{|p_z|}{|\p|}\right)
  &= - \frac{4\pi m_B}{m_{AB}}
  \int_{-\infty}^\infty\!\frac{d\k dk_z}{(2\pi)^3}
  \frac{\cos\bigl(l\theta_{\hat{\k}\cdot\hat{\p}}\bigr)}
  {\frac{m_B}{m_A}(\k+\p)^2+\k^2+k_z^{\,2}+\p^2+p_z^{\,2}}
  \frac{|\k|^{\gamma_l+1}}{\sqrt{\frac{m_B}{M}\k^2+\k^2+k_z^{\,2}}}
  \,S_l\!\left(\frac{|k_z|}{|\k|}\right).
 \end{split}
\end{equation}
Rescalings of the variables $\k\to\sqrt{\frac{m_A}M}\k$ and
$\p\to\sqrt{\frac{m_A}M}\p$ and redefinition of the unknown function
$S_l(|p_z|/|\p|)$ lead to the result shown in
Ref.~\cite{Nishida:2008kr}.  The scaling dimension of the renormalized
composite operator $\O_{ABB}^{(l)}$ is given by
\begin{equation}
 \Delta_{ABB}^{(l)} 
  = \Delta_{\phi}+\Delta_{\phi_B}+l+\left(\gamma_l+1-l\right)
  = 4+\gamma_l.
\end{equation}

By solving the integral equation (\ref{eq:gamma_ABB}) numerically, we
find that the anomalous dimension $\gamma_1$ in the $p$-wave channel
$l=1$ decreases with decreasing the mass ratio $m_A/m_B$ and eventually
becomes complex as $\gamma_1=-2\pm i\,\mathrm{Im}\,(\gamma_1)$ when
$m_A/m_B<0.0351287$~\cite{Nishida:2008kr}.  This implies the existence
of the Efimov bound states in the $p$-wave $ABB$ three-body system [see
discussions about Eqs.~(\ref{eq:efimov}) and (\ref{eq:discrete})].
Furthermore, in the range of mass ratio
$0.0351287<u<0.0660841$~\cite{Nishida:2008kr}, the anomalous dimension
is $-2<\gamma_1<-1$ and thus the scaling dimension of the three-body
composite operator $\O_{ABB}^{(l=1)}$ becomes
$2<\Delta_{ABB}^{(l=1)}<3$.  As a consequence, an additional $ABB$
three-body resonance can be introduced and the resulting system provides
a novel nonrelativistic defect CFT describing two-species fermions with
both two-body ($AB$) and three-body ($ABB$) resonances in the 2D-3D
mixture [see discussions about Eq.~(\ref{eq:resonance})].

It would be difficult to determine scaling dimensions of composite
operators with particles more than three.  However, it is possible to
estimate them by numerically solving the energy eigenvalue problems of
$H_\mathrm{osc}$ with the help of the operator-state correspondence
(\ref{eq:correspondence}) or by using the analytic $\epsilon$ expansions
around the special dimensions $d_B\to4$ and $d_B\to2$ (see
Sec.~\ref{sec:epsilon})~\cite{Nishida:2007pj}.

\subsection{Operator product expansions}

\begin{figure}[tp]
 \includegraphics[width=0.56\textwidth,clip]{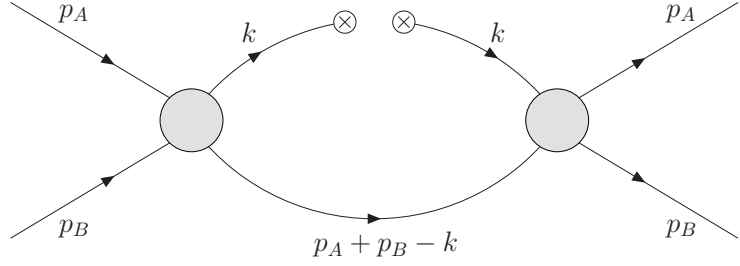}
 \caption{Feynman diagram evaluated in Eqs.~(\ref{eq:element_A}) and
 (\ref{eq:element_B}).  The solid lines are the propagators of
 $\psi_A$ and $\psi_B$ fields and the shaded bulbs represent the
 two-particle scattering amplitude $i\mathcal{A}(p_A+p_B)$.
 \label{fig:ope}}
\end{figure}

Here we consider an arbitrary effective scattering length
$-\infty<a_\eff^{-1}<\infty$ and study operator product expansions
(OPEs) in our defect quantum field theory (\ref{eq:action}).  We first
work on the following OPE:
\begin{equation}\label{eq:OPE_A}
 \psi_A^\+\!\left(\x-\frac\y2\right)
  \psi_A\!\left(\x+\frac\y2\right)
  = \sum_nW_{A,n}(\y)\O_n(\x).
\end{equation}
Here $W_{A,n}(\y)$ are Wilson coefficients and $\O_n(\x)$ are
renormalized defect operators.  We remind that $\x$ and $\y$ are
two-dimensional coordinates on the defect.  We can determine the lowest
three $W_{A,n}$ and $\O_n$ by evaluating the matrix elements of the
both sides of Eq.~(\ref{eq:OPE_A}) between two-particle states
$\<p_A,p_B|$ and $|p_A,p_B\>$.  According to
Ref.~\cite{Braaten:2008uh}, we shall consider the Feynman diagram
depicted in Fig.~\ref{fig:ope} that has nonanalyticity at $|\y|=0$.  By
denoting the total energy and momentum as $p=p_A+p_B$, the matrix
element of the left-hand side of Eq.~(\ref{eq:OPE_A}) becomes
\begin{equation}\label{eq:element_A}
 \begin{split}
  \left\<\psi_A^\+\!\left(\x-\frac\y2\right)
  \psi_A\!\left(\x+\frac\y2\right)\right\>_\mathrm{fig:\ref{fig:ope}}
  &= [i\mathcal{A}(p)]^2\int\!\frac{dk_0d\k}{(2\pi)^3}
  \,e^{i\k\cdot\y}\,iG_A(k)iG_A(k)iG_B(p-k) \\
  &= \frac{m_Bm_{AB}}{2\pi\sqrt{-2m_B\mathcal{E}_p}}\mathcal{A}(p)^2
  e^{i\frac{m_A}{M}\p\cdot\y-|\y|\sqrt{-2m_{AB}\mathcal{E}_p}}.
 \end{split}
\end{equation}
Here $\mathcal{A}(p)$ is the two-particle scattering amplitude given in
Eq.~(\ref{eq:T-matrix}) and we introduced a shorthand notation
$\mathcal{E}_p\equiv p_0-\frac{\p^2}{2M}+i0^+$.  If we expand the
exponential in terms of $|\y|$, the terms with odd powers of $|\y|$ are
nonanalytic at $|\y|=0$:
\begin{equation}\label{eq:AA}
 \begin{split}
  \left\<\psi_A^\+\!\left(\x-\frac\y2\right)
  \psi_A\!\left(\x+\frac\y2\right)\right\>_\mathrm{fig:\ref{fig:ope}}
  &= \frac{m_Bm_{AB}}{2\pi\sqrt{-2m_B\mathcal{E}_p}}
  \mathcal{A}(p)^2e^{i\frac{m_A}{M}\p\cdot\y}
  - \frac{m_{AB}\sqrt{m_Bm_{AB}}}{2\pi}\mathcal{A}(p)^2|\y|
  + O(\y^2).
 \end{split}
\end{equation}
The first term expanded in powers of $\y$ can be easily identified with
the Taylor series of the left-hand side:
\begin{equation}
 \frac{m_Bm_{AB}}{2\pi\sqrt{-2m_B\mathcal{E}_p}}
  \mathcal{A}(p)^2e^{i\frac{m_A}{M}\p\cdot\y}
  = \<\psi_A^\+\psi_A(\x)\>_\mathrm{fig:\ref{fig:ope}}
  + \frac{\y}2\cdot\<\psi_A^\+\tensor{\grad}\psi_A(\x)\>_\mathrm{fig:\ref{fig:ope}}
  + \cdots.
\end{equation}
Below we will show that the second term in Eq.~(\ref{eq:AA}) can be
identified with the matrix element of the defect operator;
$\psi_A^\+\psi_B^\+\psi_B\psi_A(\x)\equiv\psi_A^\+(\x)\psi_B^\+(\x,0)\psi_B(\x,0)\psi_A(\x)$.

The matrix element of $\psi_A^\+\psi_B^\+\psi_B\psi_A(\x)$ between the
same two-particle states $\<p_A,p_B|$ and $|p_A,p_B\>$ is evaluated as
\begin{equation}\label{eq:ABBA}
 \<\psi_A^\+\psi_B^\+\psi_B\psi_A(\x)\>
  = \left[1 + i\mathcal{A}(p)
  \int\!\frac{dk_0d\k}{(2\pi)^3}\,iG_A(p-k)iG_B(k)\right]^2 \\
  = \frac{\mathcal{A}(p)^2}{g_0^{\,2}},
\end{equation}
where we used Eq.~(\ref{eq:one-loop}).  By comparing Eq.~(\ref{eq:ABBA})
with the second term in Eq.~(\ref{eq:AA}), we find the OPE of
$\psi_A^\+\!\left(\x-\frac\y2\right)\psi_A\!\left(\x+\frac\y2\right)$ to
be
\begin{equation}
 \psi_A^\+\!\left(\x-\frac\y2\right)
  \psi_A\!\left(\x+\frac\y2\right)
  = \psi_A^\+\psi_A(\x)
  + \frac{\y}2\cdot\psi_A^\+\tensor{\grad}\psi_A(\x)
  - \frac{m_{AB}\sqrt{m_Bm_{AB}}}{2\pi}|\y|\,
  g_0^{\,2}\psi_A^\+\psi_B^\+\psi_B\psi_A(\x) + O(\y^2).
\end{equation}
Here $g_0^{\,2}\psi_A^\+\psi_B^\+\psi_B\psi_A(\x)$ is the renormalized
defect operator having finite matrix elements.  This result is a
generalization of the OPE studied in the usual 3D case in
Ref.~\cite{Braaten:2008uh} to our 2D-3D mixture.  In particular, the
existence of the nonanalytic term in $|\y|$ implies that the
two-dimensional momentum distribution of $A$ atoms has the following
large-momentum tail:
\begin{equation}
 \begin{split}
  \rho_A(|\k|) &= \int\!d\y\,e^{-i\k\cdot\y}
  \left\<\psi_A^\+\!\left(\x-\frac\y2\right)
  \psi_A\!\left(\x+\frac\y2\right)\right\>_\text{any} \\
  &\to \frac{m_{AB}\sqrt{m_Bm_{AB}}}{|\k|^3}
  \<g_0^{\,2}\psi_A^\+\psi_B^\+\psi_B\psi_A(\x)\>_\text{any}
  \qquad(|\k|\to\infty).
 \end{split}
\end{equation}
Here the expectation value can be taken with any state in the system,
for example, at finite densities of $A$ and $B$ atoms and at finite
temperature.  The quantity in the right-hand side is called the contact
density and given by
$\<g_0^{\,2}\psi_A^\+\psi_B^\+\psi_B\psi_A(\x)\>_\mathrm{any}\to4\pi^2a_\eff^{\,2}\,n_An_B/(m_Bm_{AB})$
in the weak coupling limit $a_\eff\to-0$.  The coefficient of the
large-momentum tail has played an important role in the exact analysis
of the unitary Fermi gas in pure 3D~\cite{Tan:2005}.  It is an important
future problem to investigate exact relationships in our 2D-3D mixed
dimensions.

The following OPE will be more interesting because it involves the
$z$-direction perpendicular to the 2D defect:
\begin{equation}\label{eq:OPE_B}
 \psi_B^\+\!\left(\x-\frac\y2,z\right)
  \psi_B\!\left(\x+\frac\y2,z\right)
  = \sum_nW_{B,n}(\y,z)\O_n(\x).
\end{equation}
Wilson coefficients $W_{B,n}(\y,z)$ and renormalized defect operators
$\O_n(\x)$ can be determined by evaluating the matrix elements of the
both sides of Eq.~(\ref{eq:OPE_B}) between the two-particle states
$\<p_A,p_B|$ and $|p_A,p_B\>$.  We shall consider the Feynman diagram
depicted in Fig.~\ref{fig:ope} again that has nonanalyticity at
$|\y|=|z|=0$.  The matrix element of the left-hand side of
Eq.~(\ref{eq:OPE_B}) becomes
\begin{equation}\label{eq:element_B}
 \begin{split}
  \left\<\psi_B^\+\!\left(\x-\frac\y2,z\right)
  \psi_B\!\left(\x+\frac\y2,z\right)\right\>_\mathrm{fig:\ref{fig:ope}}
  &= [i\mathcal{A}(p)]^2\int\!\frac{dk_0d\k}{(2\pi)^3}
  \,e^{i\k\cdot\y}\,iG_B(k;z)iG_B(k;-z)iG_A(p-k) \\
  &= m_Bm_{AB}\mathcal{A}(p)^2 e^{i\frac{m_B}{M}\p\cdot\y}
  \int\!\frac{d\k}{(2\pi)^2}\frac{e^{i\k\cdot\y-2|z|
  \sqrt{\frac{m_B}{m_{AB}}\k^2-2m_B\mathcal{E}_p}}}
  {\k^2-2m_{AB}\mathcal{E}_p}.
 \end{split}
\end{equation}
When $|z|\neq0$ is fixed, the right-hand side is analytic in terms of
$\y$ and therefore the OPE of
$\psi_B^\+\!\left(\x-\frac\y2,z\right)\psi_B\!\left(\x+\frac\y2,z\right)$
is simply given by its Taylor series in powers of $\y$.  This is natural
because there is no interaction with $\psi_A(\x)$ away from the 2D
defect located at $z=0$.

We now set $\y=\0$ and study the OPE of
$\psi_B^\+\psi_B(\x,z)\equiv\psi_B^\+(\x,z)\psi_B(\x,z)$ as a function
of the distance from the 2D defect $|z|$ (termed defect operator product
expansion).  Performing the integration over $\k$ in
Eq.~(\ref{eq:element_B}) with $\y=\0$, we obtain
\begin{equation}
 \begin{split}
  \<\psi_B^\+\psi_B(\x,z)\>_\mathrm{fig:\ref{fig:ope}}
  &= -\frac{m_Bm_{AB}}{2\pi}\mathcal{A}(p)^2
  \,\mathrm{Ei}\!\left(-2|z|\sqrt{-2m_B\mathcal{E}_p}\right) \\
  &= -\frac{m_Bm_{AB}}{2\pi}\mathcal{A}(p)^2
  \ln\!\left(2\,e^{\gamma_\mathrm{E}}|z|\sqrt{-2m_B\mathcal{E}_p}\right)
  + O(|z|),
 \end{split}
\end{equation}
where $\mathrm{Ei}(x)\equiv\int_{-x}^\infty\!\frac{dt}{t}e^{-t}$.  
We can identify the lowest order term in the right-hand side with
\begin{equation}
 -\frac{m_Bm_{AB}}{2\pi}\mathcal{A}(p)^2
  \ln\!\left(2\,e^{\gamma_\mathrm{E}}|z|\sqrt{-2m_B\mathcal{E}_p}\right)
  = \<\psi_B^\+\psi_B(\x,0)\>^{(\lambda)}_\mathrm{fig:\ref{fig:ope}}
  -\frac{m_Bm_{AB}}{2\pi}\ln\left(2\,e^{\gamma_\mathrm{E}}|z|\lambda\right)
  \<g_0^{\,2}\psi_A^\+\psi_B^\+\psi_B\psi_A(\x)\>,
\end{equation}
where $\lambda$ is an arbitrary momentum scale.  Therefore we find the
defect OPE of $\psi_B^\+\psi_B(\x,z)$ to be
\begin{equation}
 \psi_B^\+\psi_B(\x,z) = \psi_B^\+\psi_B(\x,0)^{(\lambda)}
  -\frac{m_Bm_{AB}}{2\pi}\ln\left(2\,e^{\gamma_\mathrm{E}}|z|\lambda\right)
  \,g_0^{\,2}\psi_A^\+\psi_B^\+\psi_B\psi_A(\x) + O(|z|).
\end{equation}
Because $\psi_B^\+\psi_B(\x,z)$ is the density operator of $B$ atoms,
the above result suggests that the density of $B$ atoms diverges
logarithmically toward the 2D defect $|z|\to0$:
\begin{equation}
 \tilde n_B(|z|) = \<\psi_B^\+\psi_B(\x,z)\>_\mathrm{any}
  \to -\frac{m_Bm_{AB}}{2\pi}\ln|z|\,
  \<g_0^{\,2}\psi_A^\+\psi_B^\+\psi_B\psi_A(\x)\>_\mathrm{any}.
 \end{equation}
The coefficient of the divergence is given by the contact density up to
the mass-dependent factor.  Further analysis to elucidate this aspect
will be worthwhile.

\subsection{Conclusion}
Two-species fermions in the 2D-3D mixed dimensions in the unitarity
limit can be regarded as a nonrelativistic defect CFT.  We derived the
reduced Schr\"odinger algebra and the operator-state correspondence in
general nonrelativistic defect CFTs.  We also studied scaling dimensions
of few-body composite operators and operator product expansions in our
2D-3D mixture.  In particular, for the stability of the many-body system
near the unitarity limit, we showed that the mass ratio has to be in the
range $0.0351287<m_\mathrm{2D}/m_\mathrm{3D}<6.35111$ to avoid the
Efimov effect~\cite{Nishida:2008kr}.  Finally, we emphasize that all
field-theoretical methods presented here to determine scaling dimensions
and critical mass ratios are widely applicable to both fermionic and
bosonic systems and also in the 1D-3D mixture~\cite{Nishida:2008kr} and
in the usual 3D case~\cite{Nishida:2007pj,Nishida:2007mr,Mehen:2007dn}.

\end{document}